# Synthesis of clathrate cerium superhydride $CeH_9$ below 100 GPa with atomic hydrogen sublattice

Nilesh P. Salke[1], M. Mahdi Davari Esfahani[2], Youjun Zhang[3,1], Ivan A. Kruglov[4,5], Jianshi Zhou[6], Yaguo Wang[6], Eran Greenberg[7], Vitali B. Prakapenka[7], Artem R. Oganov[8,4,9,*], Jung-Fu Lin[10,*]

[1]Center for High Pressure Science & Technology Advanced Research (HPSTAR), Shanghai, 201203, China
[2]Department of Geosciences, Center for Materials by Design, and Institute for Advanced Computational Science, State University of New York, Stony Brook, New York 11794-2100, USA
[3]Institute of Atomic and Molecular Physics, Sichuan University, Chengdu 610065, China
[4]Department of Problems of Physics and Energetics, Moscow Institute of Physics and Technology, 9 Institutskiy Lane, Dolgoprudny City, Moscow Region 141700, Russia
[5]Dukhov Research Institute of Automatics (VNIIA), Moscow 127055, Russia
[6]Department of Mechanical Engineering, The University of Texas at Austin, Austin, Texas 78712, USA
[7]Center for Advanced Radiation Sources, University of Chicago, Illinois, 60637, USA
[8]Skolkovo Institute of Science and Technology, Skolkovo Innovation Center, 3 Nobel Street, Moscow 143026, Russia
[9]International Center for Materials Design, Northwestern Polytechnical University, Xi'an 710072, China
[10]Department of Geological Sciences, The University of Texas at Austin, Austin, Texas 78712, USA

**Corresponding author: Jung-Fu Lin* (afu@jsg.utexas.edu)
**Corresponding author: Artem R. Oganov* (artem.oganov@stonybrook.edu)

## Abstract

Hydrogen-rich superhydrides are believed to be very promising high-$T_c$ superconductors as they are expected to mimic characteristics of metallic hydrogen. Recent experiments discovered superhydrides at very high pressures, e.g. $FeH_5$ at 130 GPa and $LaH_{10}$ at 170 GPa. With the motivation of discovering new hydrogen-rich high-$T_c$ superconductors at lowest possible pressure, here we report the prediction and experimental synthesis of cerium superhydride $CeH_9$ below 100 GPa in the laser-heated diamond anvil cell. Ab-initio calculations were carried to evaluate the detailed chemistry of the Ce-H system and to understand the structure, stability and superconductivity of $CeH_9$. $CeH_9$ crystallizes in a $P6_3/mmc$ clathrate structure with a substantially dense 3-dimensional hydrogen sublattice at 100 GPa. These findings shed a new light on the search for superhydrides in close proximity with atomic hydrogen within a feasible pressure range. Discovery of superhydride $CeH_9$ provides a practical platform to further investigate and understand conventional superconductivity in hydrogen rich superhydrides.

## Main text

Metallization of hydrogen under high pressure has been a topic of great scientific interest in the past few decades mainly due to expectations of room-temperature superconductivity[1-7]. Hydrogen is expected to become metallic under high pressure above 400 GPa[7-9]. But achieving such pressures is very challenging in diamond anvil cell experiments, mainly due to diamond failure and lack of a reliable probe on the tiny sample volumes at such high pressures. Meanwhile hydrogen-rich hydrides are also expected to achieve high-$T_c$ superconductivity perhaps at a much lower pressure than that of required for metallic hydrogen[10-12]. Hydrides and metallic hydrogen are expected to be conventional superconductors due to the presence of hydrogen. For conventional superconductivity, high phonon frequency, strong electron-phonon coupling and high density of states at the Fermi level are the essential conditions for superconductivity with Cooper pairs mediated by electron-phonon interaction[13]. Existence of hydrogen in hydride sublattice satisfies all these conditions as the low mass of hydrogen results in high phonon frequency, covalent bonding is favourable for strong electron-phonon coupling, and metallization under high pressure can result in high electronic density of states at the Fermi level[10]. Within this view the remarkable prediction and experimental confirmation of superconductivity at a record high $T_c$ of 203 K under pressure of 150 GPa in $H_3S$ makes sense[14,15]. The discovery of superconductivity in $H_3S$ has given hopes to achieve room-temperature superconductivity in hydrogen-rich systems under high pressure.

Hydrogen readily reacts with most elements to form binary hydrides.[16,17] Several hydrogenic motifs such as $H^{\delta-}$, $H_2^{\delta-}$, $H_3^-$, $H_3^+$, $H_4^-$ and $H_5^+$, and infinite chains, layers, frameworks were predicted to occur in high-pressure hydrides[18-20]. 1-D hydrogen chains and 3-D clathrate structures with hydrogen cage were predicted and found to be good candidates for high $T_c$ superconductivity[18]. $H_3S$ with highest $T_c$ of 203 K has body centered cubic structure which can also be visualized as sulphur atom surrounded by 3-dimensional hydrogen cage. Hydrides possessing $H_2$-units are not prone to have high $T_c$ as they tend to have low densities of states at the Fermi level[21]. Recent theoretical predictions have reported several systems with unusually high hydrogen content, termed as polyhydrides/superhydrides, to become stable under high pressure and to have very high $T_c$ under pressure[20-27]. Notably $CaH_6$[22], $MgH_6$[24], $YH_6$[25], $YH_9$[21], $YH_{10}$[21,26], $LaH_{10}$[26], $AcH_{10}$ and $AcH_{16}$ [27] were predicted to have $T_c$ above 235 K. Most of the phases with $T_c$ close to room temperature are predicted to have a clathrate structure with hydrogen forming a cage around metal atom ($M$). In $MH_6$, $MH_9$ and $MH_{10}$ compounds, metal atoms are located within $H_{24}$, $H_{29}$ and $H_{32}$ cages respectively[21,22,25,26]. However, it is essential to know the experimental pressure-temperature condition to stabilize a hydride before carrying out the further electrical or magnetic measurement to verify the superconductivity. Recently, a handful of experiments were reported to synthesize new superhydrides under pressure, particularly $FeH_5$ at 130 GPa[28], $LaH_{10}$ at 170 GPa[29], $UH_7$, $UH_8$ and $UH_9$ above 37 GPa[30]. There were also experimental reports about synthesis of new and unusual hydrides under pressure, such as $LiH_6$[31], $NaH_7$[32], $Xe(H_2)_7$[33] and $HI(H_2)_{13}$[34] with $H_2$-like molecular units. Synthesis of $FeH_5$ and $LaH_{10}$ without any $H_2$-like unit is very intriguing. Following to the experimental synthesis of $LaH_{10}$, recently the experimental $T_c$ of 260 K at 190 GPa and 215 K at 150 GPa is claimed for $LaH_{10}$ by two different research groups by electrical conductivity measurement [35,36], if proven it would be a new record. However, there is a huge discrepancy in the claimed $T_c$ by two independent measurement, mainly because the very high pressure of 170 GPa involved in the synthesis of $LaH_{10}$. At such pressure the verification of $T_c$ becomes challenging task. Interestingly, $FeH_5$ with layered structured consisted 2-dimensional atomic hydrogen slabs. Also, nearest H-H distance in $FeH_5$ was reported to be ~1.336 Å at ~100 GPa[28], whereas for $LaH_{10}$ it was ~1.196 Å at ~120 GPa[29]. $LaH_{10}$ was claimed as closest analogue to solid atomic metallic hydrogen based on nearest H-H distance.[29] But the pressure required to stabilize $FeH_5$ and $LaH_{10}$ phases was 130 and 170 GPa, respectively[28,29], which is relatively high. It would be desirable to get hydrogen-rich phases with the lowest possible pressure for further experimental verifications and realistic technological application. Synthesis of superhydrides at lower pressures would give an opportunity to further investigate the nature of superconductivity and atomic hydrogen by other techniques for an in-depth understanding. Studies on the synthesis path and structure of superhydrides also help to build a deeper understanding of hydride chemistry. Besides superconductivity, hydrides are also very important as hydrogen storage materials for next generation energy related applications[37]. Recently, Peng et al. (2017) predicted that hydrogen-rich $CeH_9$ with $P6_3/mmc$ structure becomes stable at a relatively low pressure of 100 GPa[21], which by itself is very interesting although their estimated superconducting $T_c$ was relatively low, < 56 K. We have carefully studied the Ce-H system in order to understand the crystal chemistry and to seek for superconductivity with, much higher $T_c$ values.

Here we report the successful synthesis of cerium superhydride $CeH_9$ below 100 GPa. Using evolutionary variable-composition searches, whole compositional space of the Ce-H system explored in a single simulation. We predicted phase stability and superconducting properties of high-pressure cerium superhydrides. Rich chemistry of cerium hydrides manifests itself in numerous stable compounds, including the experimentally synthesized $CeH_3$ and superhydrides $CeH_9$. We have carried out a direct elemental reaction between cerium and hydrogen using a laser-heated diamond anvil cell (DAC) coupled with synchrotron x-ray diffraction (XRD). It is found that heating plays an essential role in the formation of Ce-H phases at high pressures. Analysis of XRD results in combination with ab initio calculations shows that $CeH_9$ crystallizes in a clathrate structure with space group $P6_3/mmc$ above 80 GPa. Each cerium atom is enclosed within a cage of $H_{29}$ in which hydrogen atoms are bonded covalently. Besides this, a previously unknown $Pm\bar{3}n$ structured $CeH_3$ ($\beta$-$UH_3$ type[38]) was synthesized at 36 GPa with laser heating. The detailed first-principles investigation of stability, structural, electronic and superconducting properties of experimentally synthesized hydrogen-rich phase was carried out. We studied, specifically electron-phonon interaction of $P6_3/mmc$-$CeH_9$ and predict that the $CeH_9$ is a high temperature superconductor with $T_c$ = 117 K at 200 GPa.

## Results

**Synthesis of various Ce-H phases.** In our experiment, various phases of the Ce-H system such as $CeH_x$ (X = 2, 2.5, 3 and 9) were synthesized successfully at high pressures. Initially, the cerium sample and hydrogen gas were loaded into the sample chamber of the DAC and were kept at 9 GPa. A small piece of gold also was loaded along with sample to calibrate pressure. At 9 GPa and ambient temperature, we found that cerium and hydrogen reacted in the sample chamber which resulted in the formation of a cerium hydride compound as shown by the XRD pattern in Fig. 1a. The corresponding XRD image is shown in supplementary Fig. S1a. All the peaks observed at 9 GPa could be indexed with the $Fm\bar{3}m$ phase of $CeH_2$. The $Fm\bar{3}m$ phase of $CeH_2$ persisted up to 33 GPa (supplementary Fig. S1b). Lebail refinements were carried to extract lattice parameters of the $CeH_2$ phase (supplementary Fig. S1c). The lattice parameters of $CeH_2$ at 9 GPa and 33 GPa were determined as $a$ = 5.370(1) and $a$ = 5.011(2) Å respectively. Pressure dependence of the unit cell volume of $CeH_2$ was fitted with a third-order Birch-Murnaghan equation of state (EOS) which yielded the fitting parameters such as unit cell volume at zero pressure $V_0$ = 44(1) Å$^3$/f.u., bulk modulus $K_0$ = 45(6) GPa and first pressure derivative of bulk modulus $K_0{'}$ = 4 (fixed) (supplementary Fig. S1d). Pressurization of $CeH_2$ up to 33 GPa did not result in any changes in crystal structure. However, microsecond pulsed laser heating of ~2000 K carried at 33 GPa resulted in obvious structural changes (supplementary Fig. S2a). The XRD pattern at 36 GPa obtained after laser heating is shown in Fig. 1b, and the corresponding XRD image is shown in supplementary Fig. S2b. The integrated XRD pattern at 36 GPa was found to be of cubic $CeH_3$ with $Pm\bar{3}n$ isomorphous to $\beta$-$UH_3$ ($\beta$-$Pm\bar{3}n$) [38]. This high-pressure phase of $CeH_3$ with $\beta$-$Pm\bar{3}n$ structure has also been predicted in our evolutionary searches to be the energetically favourable phase below 10 GPa and is 32 meV/atom higher in enthalpy than the most favourable $CeH_3$ at the synthesized pressure of 36 GPa (supplementary Fig. S3). $\beta$-$UH_3$ type $Pm\bar{3}n$-$CeH_3$ is being reported for the first time here. The experimental lattice parameters of the $\beta$-$Pm\bar{3}n$ phase at 36 GPa are $a$ = 6.2788(3) Å. In $\beta$-$Pm\bar{3}n$ structure of $CeH_3$, cerium atoms occupy 2a (0, 0, 0) and 6c (1/4, 0, 1/2) Wyckoff positions[38]. Unfortunately, very low x-ray scattering factor of hydrogen atom did not allow us to determine the exact position of hydrogen atoms in $CeH_3$ unit cell from the experimental XRD data. Theoretical calculations yielded the Wyckoff position for the hydrogen atoms as 24K (0, 0.1580, 0.6935) at 35 GPa with lattice parameter $a$ = 6.2471 Å, which is highly consistent with the experimental value of $\beta$-$Pm\bar{3}n$-$CeH_3$ observed at 36 GPa. The $\beta$-$Pm\bar{3}n$ phase of $CeH_3$ proved stable with further compression up to 80 GPa and also sustained laser heating at an intermediate pressure of 60 GPa (supplementary Fig. S2a), which agrees with our predictions. A third-order Birch-Murnaghan EOS was used to fit the $P$-$V$ data of $CeH_3$ (see supplementary Fig. S2d), fitting parameters are $V_0$ = 39.7(4) Å$^3$/f. u., $K_0$ = 86(4) GPa and $K_0'$ = 4 (fixed).

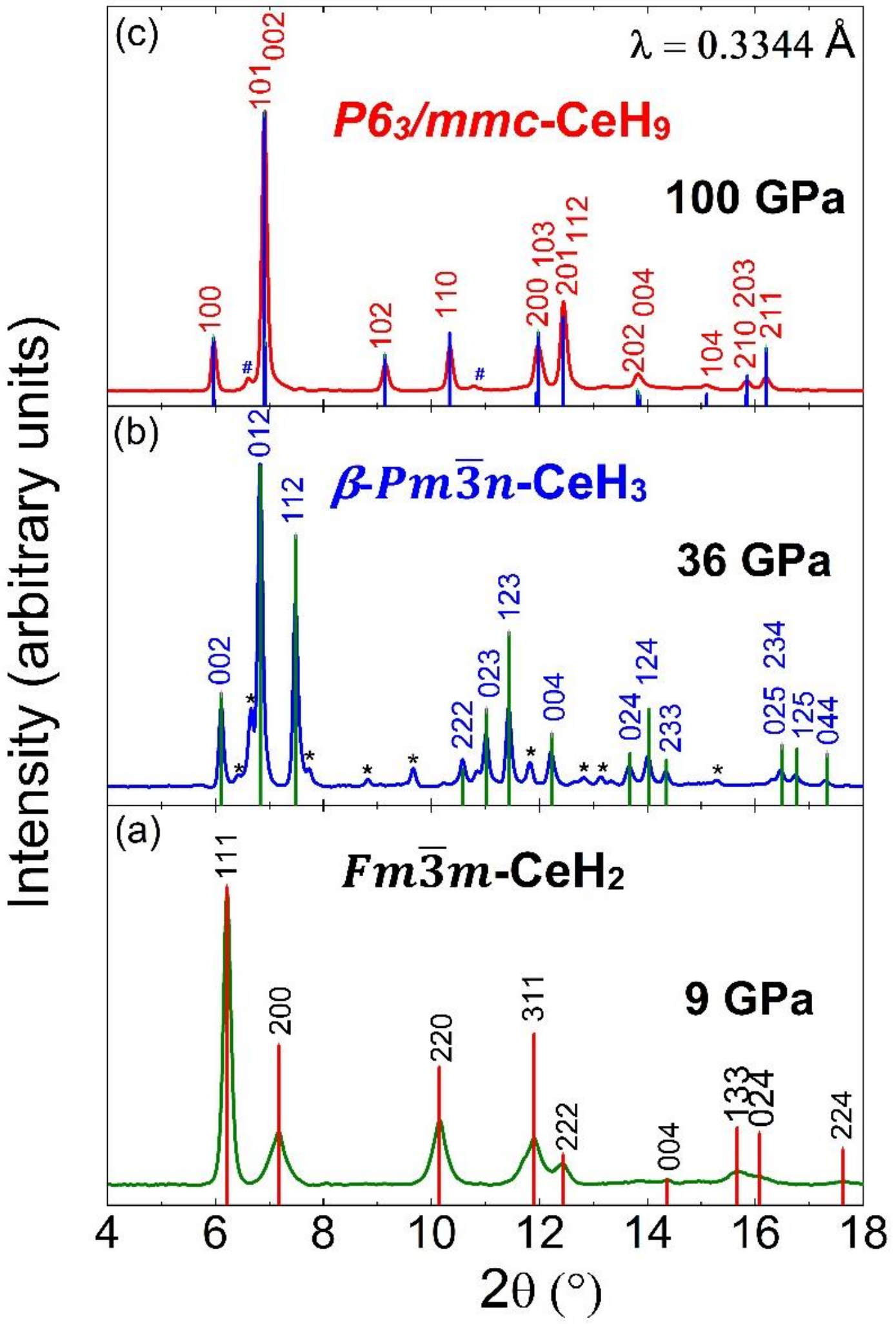

**Figure 1: Representative integrated XRD patterns of high-pressure phases in Ce-H system up to 100 GPa**. Typical XRD patterns of (a) $CeH_2$, (b) $CeH_3$, and (c) $CeH_9$ obtained at 9, 36 and 100 GPa of pressure respectively. Vertical lines indicate the indexing with calculated intensity for respective crystal structure. $CeH_2$, $CeH_3$ and $CeH_9$ crystallize in space group $Fm\overline{3}m$, $Pm\overline{3}n$ ($\beta$-$UH_3$ type) and $P6_3/mmc$ respectively. Unidentified weak peaks in (b) and (c) are marked with * and #, respectively, These additional peaks could not be identified or indexed with any of the known or predicted phases of Ce[39] or $CeH_x$[21], as well as cerium oxides[40,41].

Several cycles of pulse laser heating with 1 µs pulse width at a repetition rate of 10kHz for a total heating duration of a few seconds each cycle was used to laser heat the sample assemblage to approximately 1700 K at 80 GPa that resulted in the emergence of new peaks (supplementary Fig. S4). These new diffraction peaks were indexed to be (101) and (002) of clathrate hexagonal phase of $CeH_9$ (see supplementary Fig. S4 and Figs. 1c and 2) predicted by our evolutionary search. With further pressurization the relative intensity of the (101) and (002) peaks of $CeH_9$ phase increased (supplementary Fig. S4). Although most of the peaks of $CeH_3$ phase were present, the intensity of the $CeH_9$ peaks became prominent and increased with pressure (supplementary Fig. S4). Several further cycles of laser heating at 88 GPa (supplementary Fig. S4) and 98 GPa improved the intensity of the hexagonal $CeH_9$ phase, as shown in Fig. 1c. A Rietveld refinement plot for the $CeH_9$ phase at 100 GPa is shown in Fig. 2a with the corresponding XRD image as an inset. $CeH_9$ crystallizes in the $P6_3/mmc$ space group with lattice parameters $a$ = 3.7110(3) and $c$ = 5.5429(7) Å at 100 GPa. Cerium atoms occupy the Wyckoff position 2d (2/3, 1/3, 1/4) in a hexagonal unit cell. Theoretical calculations established positions of hydrogens to be 2b (0, 0, 1/4), 4f (1/3, 2/3, 0.1499) and 12k (0.1565, 0.8435, 0.4404) at 100 GPa. The crystal structure of $CeH_9$ is shown in Fig. 2b. The experimentally observed $P6_3/mmc$ $CeH_9$ phase and its structural parameters are perfectly consistent with our calculations. Calculated EOS parameters for $CeH_9$ phase are as follows, $V_0$ = 53.4(2) Å$^3$/ f.u., $K_0$ = 80.5(13) GPa and $K_0'$ = 4 (fixed).

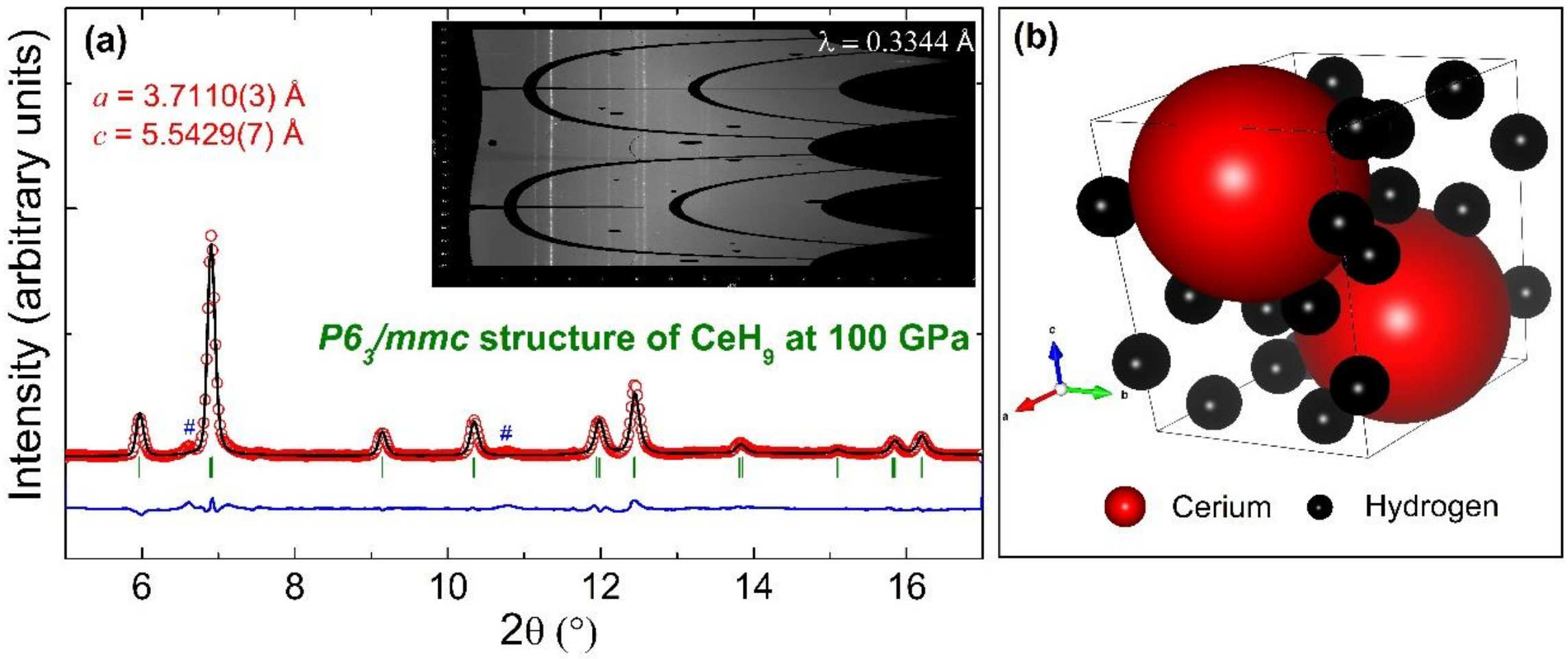

**Figure 2: Rietveld refinement of the hexagonal $CeH_9$ in $P6_3/mmc$ structure at 100 GPa**. (a) Rietveld refinement plot of powder XRD data at 100 GPa. Red open circles: experimental data of $CeH_9$ in $P6_3/mmc$ structure at 100 GPa; black line: model fit for the structure; green vertical lines: Bragg diffraction positions of the structure; blue line: residues. Reliability parameters for the Rietveld refinement are as follows (in %): $R_p$ = 14.5 $R_{wp}$ = 18.4, $R_{Bragg}$ = 8.05. # represents unidentified weak peaks. Inset shows the Pilatus XRD image of corresponding powder XRD pattern with the incident x-ray wavelength of 0.3344 A. (b) Crystal structure model of $P6_3/mmc$ structured $CeH_9$. Red and black spheres represent cerium and hydrogen atoms respectively.

During the decompression cycle, the $CeH_9$ phase was observed to become unstable at pressures below 93 GPa (supplementary Fig. S5). Laser heating was also carried out during the decompression cycle at 79 and 54 GPa. Laser heating in decompression cycle did not show any precise changes in XRD pattern. Upon further decompression, the $\beta$-$Pm\bar{3}n$-$CeH_3$ phase reappeared below 50 GPa (supplementary Fig. S5). Finally, after complete decompression, the $\beta$-$Pm\bar{3}n$-$CeH_3$ phase was recovered at ambient conditions along with tetragonal $Ce_2H_5$ (supplementary Fig. S5). The $P$-$T$ path for the formation and stability of various Ce-H phases, observed in our experiments, can be seen in Fig. S6.

**Theoretical calculations and prediction of cerium hydrides.** First-principles calculations were carried out to understand the detailed chemistry of the Ce-H system, dynamical stability, and structural and electronic band structures of experimentally synthesized superhydride phases. We performed variable-composition evolutionary searches at 0, 50, 100, 150, 200 and 250 GPa. The thermodynamic convex hull construction at different pressures is depicted in supplementary Fig S7. We predicted stable cerium hydrides and their stable structures at different pressure conditions, which are shown in the pressure-composition phase diagram in Fig. 3. Several compounds such as *I4/mmm*-$CeH_4$, *P6$_3$mc*-$CeH_6$, *P6$_3$mc*-$CeH_8$, *P6$_3$/mmc*-$CeH_9$ and $Fm\bar{3}m$-$CeH_{10}$, were predicted, in addition to finding three known compounds $CeH_2$, $Ce_2H_5$. High-pressure phase of the $CeH_3$ was also predicted with space group $Pm\bar{3}n$, as shown in Fig. 3. Because of high concentration of hydrogen in hydrogen-rich hydrides, anharmonic effects might be important in determining the relative stability of hydrogen-rich phases, however, in our previous studies[20,42], we showed that quantum effects do not change the topology of the phase diagram, and quantitative effects are just moderate shifts in transition pressures. For example, the inclusion of ZPE lowers the formation enthalpies of *Ama2* and *C2/m* structures and shifts the transition pressure *Ama2* → *C2/m* from 300 to 278 GPa[42]. Among the stable phases predicted, we have synthesized $Fm\bar{3}m$-$CeH_2$, $I4_1md$-$Ce_2H_5$ *β*-$Pm\bar{3}n$-$CeH_3$ and $P6_3/mmc$-$CeH_9$. Our pressure-composition phase diagram shows pressure ranges of stability for all the predicted phases along with experimentally known compounds. It clearly shows that higher pressures favour higher hydrogen content compounds, which is consistent with our experiment done at different pressure conditions. Previously known compounds $Fm\bar{3}m$-$CeH_2$ and $I4_1md$-$Ce_2H_5$ are predicted to be stable only below 8 and 1.5 GPa, respectively. Increase of pressure leads to the formation of *I4/mmm*-$CeH_4$ above 32 GPa. $CeH_6$ and $CeH_8$ are stable in relatively narrow pressure ranges from 26 to 68 GPa, and 55 to 95 GPa, respectively and that is probably why they are not observed in our experiment. *P6$_3$/mmc*-$CeH_9$ becomes stable at pressures above 78 GPa, which agrees with our high-pressure experiments where it was synthesized above 80 GPa. Detailed structural information on the predicted phases can be found in the supplementary Table S1. Among the predicted stable cerium hydrides, we focus our modelling on hydrogen-rich $CeH_9$, since a higher hydrogen to metal ratio in hydrides is expected to correlate with higher $T_c$ superconductivity in hydrogen-rich hydrides[11].

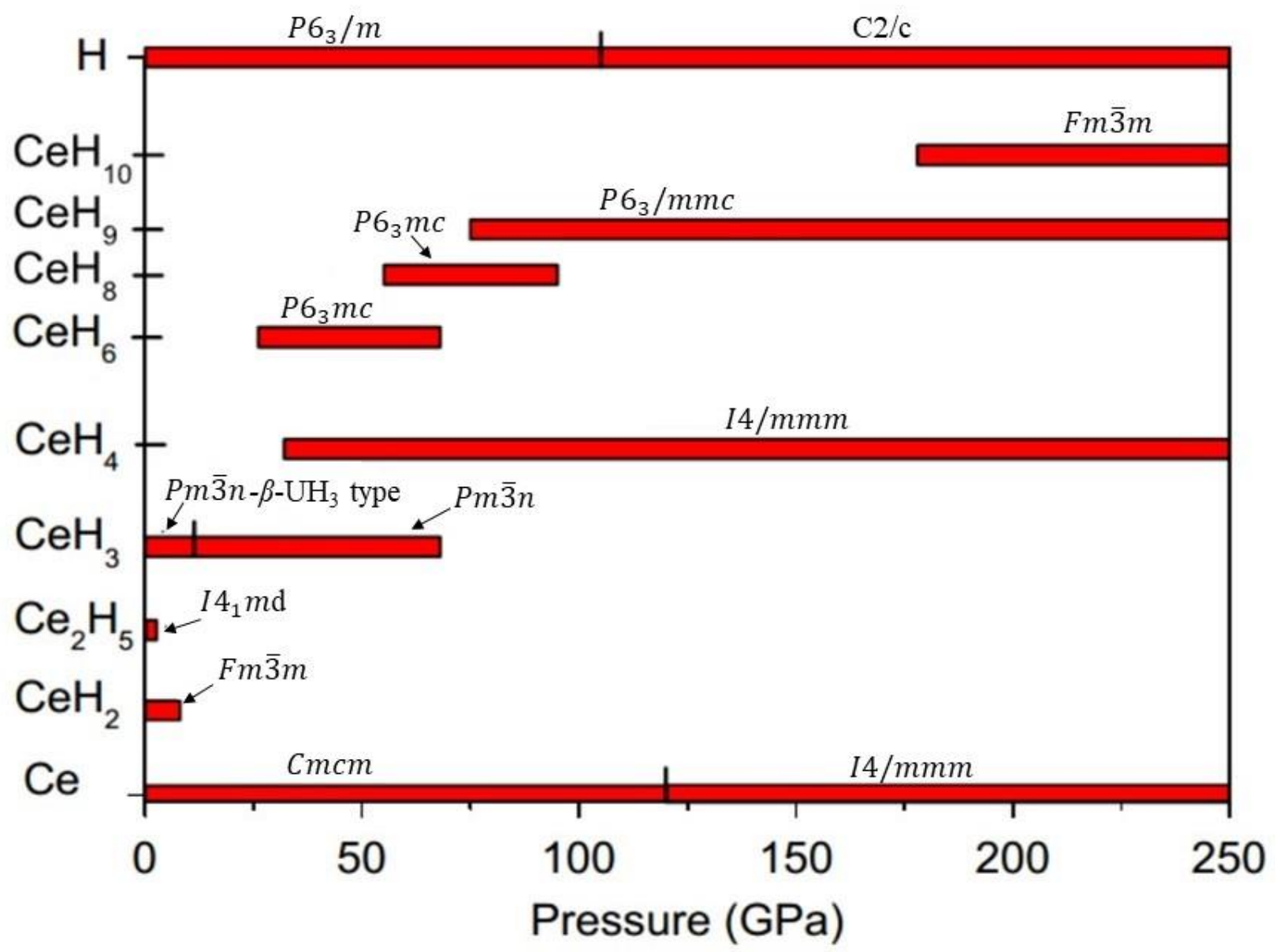

**Figure 3: Pressure-composition phase diagram of theoretically predicted stable phases in the Ce-H system at high pressures.** Red horizontal bars correspond to each Ce-H composition with a predicted crystal structure with its pressure range of stability calculated using the evolutionary structure search USPEX. The experimentally discovered *P6₃/mmc*-$CeH_9$ is predicted to be stable in the pressure range 78-250 GPa.

## Discussion

Addition of hydrogen in metallic sublattice expands the unit cell volume. In most cases, rate of increase in volume is proportional to hydrogen content in hydride. In hydrides, expansion of volume with respect to pure metal volume were frequently used to establish the hydrogen content and stoichiometry[29]. In order to ascertain the stoichiometry of high pressure superhydride phase observed above 80 GPa, we have compared the ideal mixture of Ce-$H_2$ solution with experimental volume per formula unit of $CeH_3$, $CeH_9$ and with theoretical EOS (see Fig. 4a). The curve representing ideal mixture of Ce and (7/2) $H_2$ lies well below the theoretical and experimental EOS of $P6_3/mmc$-$CeH_9$ in the pressure range 80 to 100 GPa. Whereas mixture of Ce and (8/2) $H_2$ partially overlaps with theoretical and experimental EOS of $CeH_9$. This indicates that the hexagonal phase observed above 80 GPa does not favour energetically $CeH_X$ with $x < 8$ and can decompose into its elemental constituents or hydride with $x > 8$ and hydrogen. On the other hand, the curve representing an ideal mixture of Ce and (9/2) $H_2$ lies well above the theoretical and experimental EOS of $CeH_9$ in the pressure range 80 to 100 GPa. This observation clearly indicated that $CeH_9$ can be stabilized in the pressure range mentioned. From our theoretical calculations and energetic considerations, it clearly signifies that the hexagonal phase observed above 80 GPa with laser heating has the $CeH_9$ stoichiometry.

We can also see that there is a fair agreement between experimental volume and theoretical EOS results (Fig. 4a) for $P6_3/mmc$-CeH$_9$ as well as $\beta$-$Pm\bar{3}n$-CeH$_3$.

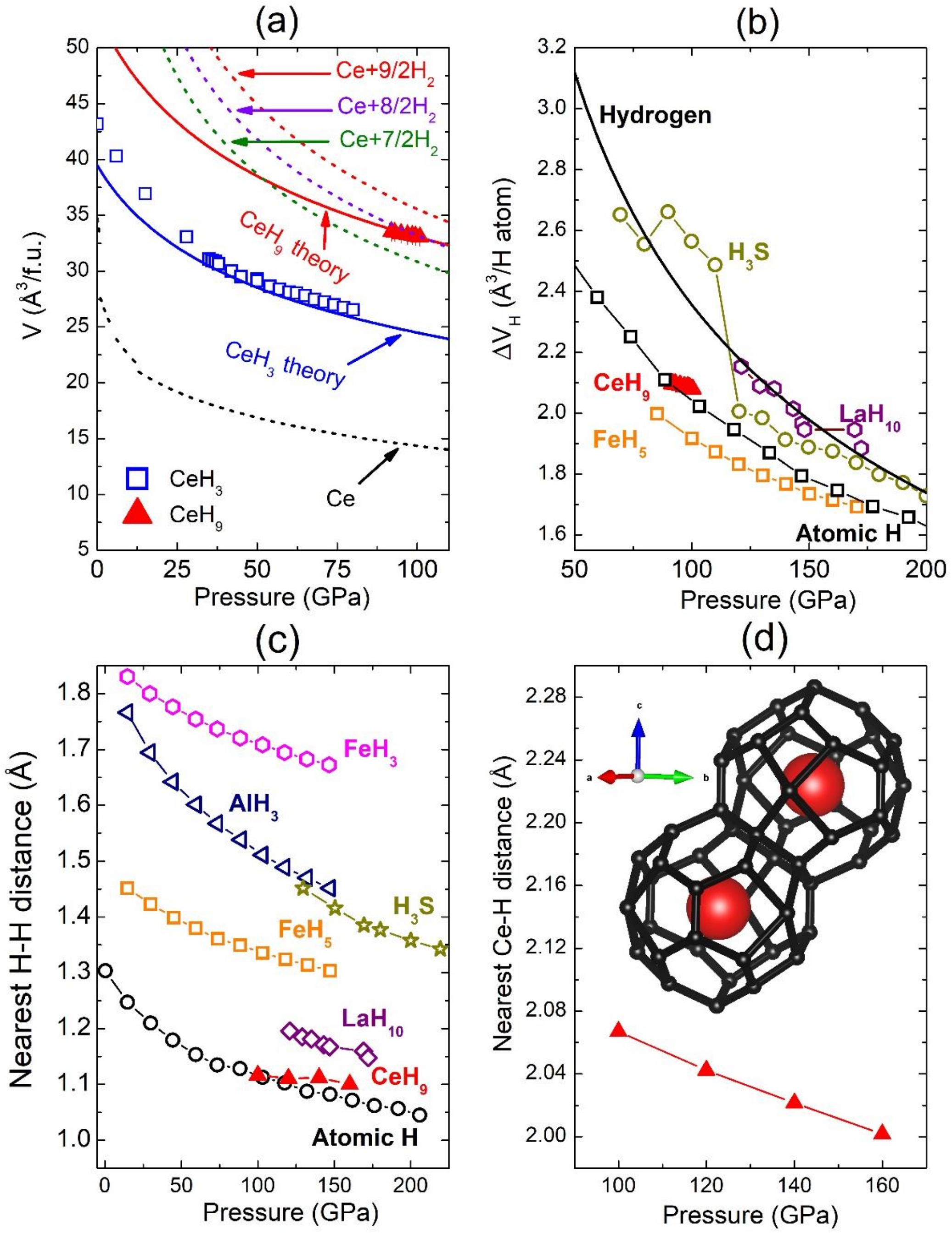

**Figure 4: Effect of hydrogen on unit cell volume and nearest neighbour distances in various hydrides at high pressures.** (a) Experimentally obtained volume per formula unit for CeH$_3$ and CeH$_9$ as a function of pressure. Theoretical EOS of CeH$_3$ and CeH$_9$ is plotted as blue and red lines, respectively. Black dashed line represents EOS of cerium metal.[39] Red, purple and green dashed curves represent ideal mixtures of Ce + (9/2)H$_2$, Ce + (8/2)H$_2$ and Ce + (7/2)H$_2$, respectively.[39,43] (b) Volume expansion per hydrogen atom ($\Delta V_H$) plotted against pressure for CeH$_9$, FeH$_5$[28], H$_3$S[14], LaH$_{10}$[29], H[43] and atomic H[28] for comparison (c) Comparison of the pressure dependence of the nearest H-H distances for CeH$_9$, FeH$_3$[44], FeH$_5$[28], AlH$_3$[45], H$_3$S[14], LaH$_{10}$[29] and atomic H[28]. (d) Nearest Ce-H distance for CeH$_9$ as a function of pressure. Inset shows Ce-H$_{29}$ clathrate cage in *P6$_3$/mmc* structure. [Pressure dependent experimental data is at 300 K, whereas theoretical data is at 0 K]

In hydrides, hydrogen is pre-compressed in the vicinity of metal atoms. Hydrogen sublattice in hydrides is expected to be identical with atomic metallic hydrogen. So, we compared the volume expansion per hydrogen atom ($\Delta V_H$) and nearest H-H distance in $CeH_9$ with simulated atomic metallic hydrogen phase extrapolated to lower pressure[28] and also with other reported hydrides as shown in Figs. 4b and 4c, respectively[28,29,44,45]. The nearest Ce-H distance in $P6_3/mmc$-$CeH_9$ is plotted in Fig. 4d. The inset of Fig. 4d shows the Ce-$H_{29}$ clathrate cage structure in which the $H_{29}$ cage is encircling a cerium atom. $\Delta V_H$ for $CeH_9$ is 2.09 Å$^3$ at 100 GPa which is lower than that for a hydrogen atom, but larger by ~8% with respect to $\Delta V_H$ of layered $FeH_5$. $\Delta V_H$ for $CeH_9$ matches with volume of atomic metallic hydrogen around 100 GPa. Formation of $CeH_9$ can also be imagined as an absolute mixing of Ce and dense atomic metallic hydrogen. It indicates that hydrogen slab surrounding Ce atom is identical to dense atomic metallic hydrogen at a lower pressure. Nearest H-H distance in clathrate $CeH_9$ is 1.116 Å at 100 GPa, which is significantly longer than the H-H bond length (0.74 Å) in $H_2$ gas molecules but is significantly lower than in other hydrides such as $AlH_3$, $FeH_3$, $FeH_5$, $H_3S$ and $LaH_{10}$ as can be seen in Fig. 4c. Surprisingly, the nearest H-H distance in $CeH_9$ almost overlaps with the H-H distance in atomic hydrogen and decreases very slowly with pressure. Among all the superhydrides, the nearest H-H distance observed in $CeH_9$ is remarkably short at 100 GPa and coincides with H-H distance of atomic metallic hydrogen. Among the reported hydrides, nearest H-H distance of $CeH_9$ is only second to the H-H distance (0.98 Å)[21] for atomic metallic hydrogen at 500 GPa at which hydrogen is in a superconducting metallic state[46]. Presence of strongly coupled hydrogen-dominant libration and stretch vibrations are the signatures of high-$T_c$ in hydrogen-rich materials[47]. Weak H-H interactions with preferred bond distances between 1.2 to 1.3 Å, the stretching and bending vibrations becomes indistinguishable, due to which all H vibrations contribute in the strong electron-phonon coupling process, eventually contributing to enhance the $T_c$ in hydrides[47]. At 100 GPa the nearest Ce-H distance is ~2.07 Å and it decreases with pressure. It is noteworthy that the clathrate structures predicted in the literature for rare earth (RE) hydrides $REH_6$, $REH_9$ and $REH_{10}$ have $H_{24}$, $H_{29}$ and $H_{30}$ cages surrounding the metal atom[21]. Among these cages, The $H_{29}$ cage has the smallest volume per formula unit for $YH_9$[21]. Clathrate $H_{29}$ cage in $CeH_9$ surrounding the Ce atom is almost 1.1 Å thick along the *a*- and *b*-axis, while it is 0.9 Å thick along *c*- axis at 100 GPa[48], whereas thickness of clathrate cage in $LaH_{10}$ is 0.9 Å[29]. Clathrate structured $CeH_9$ can be visualized as 3-dimensional atomic metallic hydrogen encapsulating Ce atoms (inset of Fig. 4d). Covalently bonded hydrogen sublattice in $CeH_9$ with bond length and $\Delta V_H$ similar to atomic metallic hydrogen is likely to have density similar to that for atomic hydrogen slab at 100 GPa. Hence $P6_3/mmc$-$CeH_9$ will be a good platform to investigate H-H properties to understand atomic metallic hydrogen. Recently, Carbotte et al, proposed a new technique to investigate superconductivity in high pressure hydrides and hydrogen based on optical properties, without four probes.[49] Superconductivity in $P6_3/mmc$-$CeH_9$ can also be evaluated using this optical technique.

Fig. 5a shows calculated electronic band structures of $CeH_9$ at 150 GPa. From Fig. 5a, it can be seen that $CeH_9$ is metallic and features numerous flat bands above the Fermi level. Noticeable density of electronic states at the Fermi level 0.73 states/eV/f.u., which is 1.4 times higher than that of previously found $H_3S$[14] and comparable to that of recently synthesized $LaH_{10}$[29] at an optimal pressure of 200 GPa, is a good sign for getting high-temperature superconductivity. (see supplementary Fig. S8 and Table S2). The main contributors to $N(E_f)$ are Ce-4f and H-1s orbitals, however, only those electrons that are coupled strongly to phonons are important. High frequency phonons are mainly related to H vibrations, owing to its light mass, which makes the largest contribution to the electron-phonon coupling constant. Analysis of electron localization function (ELF) shows a moderate ELF value 0.64 between H atoms within the unit, suggesting weak covalent interaction, which forms a 3D hydrogen network i.e., $H_{29}$ cage consisting of $H_4$, $H_5$ and $H_6$ rings. Very low ELF value between Ce and H indicates that no bonds were present between the Ce and H atoms (supplementary Fig. S9).

We performed phonon calculations in the thermodynamic stability range of $CeH_9$ i.e., above 80 GPa. Lattice dynamics calculations indicate that the $P6_3/mmc$-$CeH_9$ phase is dynamically stable at 150 GPa (Fig. 5b). Selected vibrational mode displacements of the $P6_3/mmc$-$CeH_9$ at the H- and K-point from the structure relaxed at 150 GPa (see supplementary Fig. S10). However, at lower pressures, e.g., 120 GPa, calculations show that some phonon modes become imaginary along the H- and K-points ([-1/3, 2/3, 1/2] and [-1/3, 2/3, 0], respectively) (supplementary Figs. S10). To resolve the soft modes of the lattice, we used generalized evolutionary metadynamics (GEM)[50], in which large displacements along the softest mode eigenvectors are used to equilibrate the system. This hybrid technique is implemented in USPEX and was successfully applied to boron and silicon and found numerous energetically competitive configurations[50]. We used supercells up to index 4 i.e., 80 atoms per cell. Using GEM, we found a stable structure with *C2/c* symmetry (Supplementary Fig. S11), which is a subgroup of $P6_3/mmc$ symmetry. Electron-phonon coupling (EPC) calculations revealed that $P6_3/mmc$-$CeH_9$ is a high temperature superconductor. Using the Allen-Dynes modified McMillan equation (Eq. 1), we estimated the superconducting transition temperature ($T_c$) to be 117 K at 200 GPa, when using the commonly accepted value of 0.10 for the Coulomb pseudopotential $\mu*$. In *P6₃/mmc*, the resulting electron-phonon coupling coefficient $\lambda$ is 2.30 at 200 GPa, which is higher than that of $H_3S$, $\lambda = 2.19$ at 200 GPa[14]. Since cerium atoms are heavy, the logarithmic average phonon frequency ($\omega_{log} = 740$ K) is lower compared with that of $H_3S$ ($\omega_{log} = 1335$ K), which results in a lower $T_c$ value of 117 K. The $T_c$ of $CeH_9$ has a lower value of 75 K at 100 GPa for *C2/c* structure. Logarithmic average phonon frequency $\omega_{log}$ of *C2/c* phase has a lower value 662 K. However, our results indicate that lower $T_c$ value is mainly related to the lower electron-phonon coupling coefficient $\lambda = 1.48$. In contrast, earlier report by Peng et. al. (2017) predicted a $T_c$ value of lower than 56 K at 100 GPa for $CeH_9$ phase with $P6_3/mmc$ structure[21]. However, our phonon calculations indicate instability of $P6_3/mmc$ phase below 120 GPa (see supplementary Fig. S10). So, our systematically carried studies estimate comparatively higher $T_c$ of 75 K for $CeH_9$ at 100 GPa for the dynamically stable *C2/c* structure. Furthermore, we have tabulated the $\omega_{log}$, $\lambda$ and $T_c$ value of $CeH_9$ along with recently predicted other superhydrides of La-H, Y-H, U-H, Ac-H and Th-H system for comparison, as shown in Table S3. Phonon dispersions curves, phonon density of states, the Eliashberg spectral function $\alpha^2F(\omega)$, and the EPC parameter $\lambda$ as a function of frequency are calculated and shown in supplementary Figs. S12 and S13 for *C2/c* and $P6_3/mmc$-$CeH_9$ at 100 and 200 GPa, respectively. It is known that quantum effects can impact the calculated superconducting transition temperatures, however in the case of strong anharmonic $H_3S$ (SG $Im\bar{3}m$), the inclusion of anharmonic correction, lowered the $T_c$ from its harmonic 204 K value[14] only to 194 K[51] at 200 GPa, and both are close to the reported experiment $T_c$ at 200 GPa[15], although the transition pressures shift is considerably large[14,52].

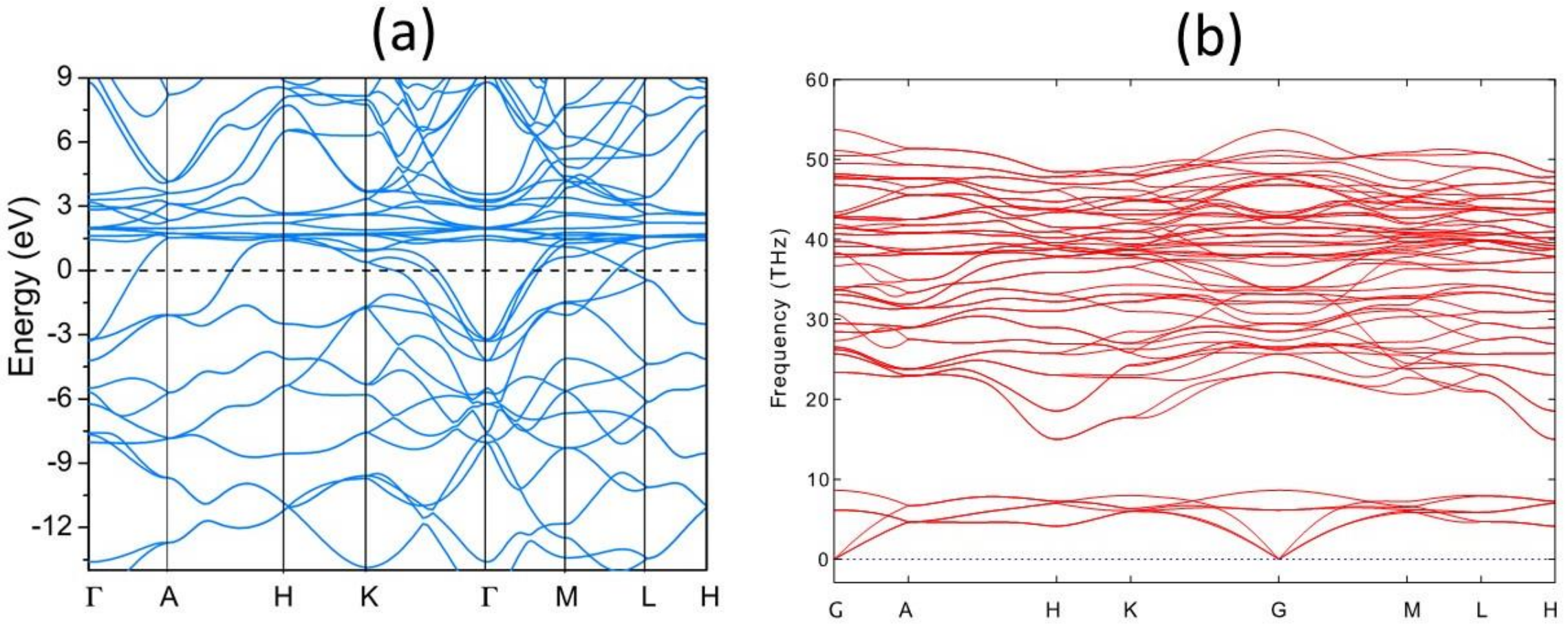

**Figure 5: Electronic band structure and phonon dispersion curves for *$P6_3/mmc$*-$CeH_9$ at 150 GPa in (a) and (b) respectively**. Dotted line in (a) indicates Fermi level. Absence of imaginary phonons in the dispersion curves Fig. (b) shows the dynamical stability of *P63/mmc*-$CeH_9$ at 150 GPa. Phonon instability at 120 GPa is shown in supplementary Fig. S10.

In summary, we have successfully synthesized a cerium superhydride phase of $CeH_9$ below 100 GPa crystallizing in the hexagonal $P6_3/mmc$ clathrate structure. In addition to this we have also synthesized a new cubic phase of $CeH_3$ with space group $Pm\bar{3}n$ ($\beta$-$UH_3$ type), which was recovered at ambient phase after complete decompression. Our studies give strong evidence for the synthesis of rare earth superhydrides and pave the way for future studies on other rare earth-hydrogen systems under extreme pressure with the aid of laser heating perhaps to make the binary hydrides $MH_x$ with x > 9. Apart from this, the estimated $T_c$ of 117 K in $P6_3/mmc$-$CeH_9$ at 200 GPa is very promising. Electron-phonon coupling in $P6_3/mmc$-$CeH_9$ is even higher than in $H_3S$ but still could not achieve higher $T_c$ due to lower logarithmic average phonon frequency. Conspicuously, the dense 3-dimensional atomic hydrogen sublattice is noted for superhydride $P6_3/mmc$-$CeH_9$ as compared with reported super/poly-hydrides and similar to atomic metallic hydrogen at 100 GPa. The discovery of $CeH_9$ at a feasible pressure range with prediction of superconductivity will certainly inspire further studies to gain in-depth understanding of hydrogen interaction in atomic hydrogen and conventional superconductivity.

## Methods

**Experimental Details.** High pressure-temperature (*P-T*) experiments were carried out using a single-sided laser heated DAC with a pair of bevelled diamond anvils of size 100-300 µm culets. Polycrystalline cerium (*Alfa Aesar*, 99.9 % purity) sample of ~5 µm thickness was loaded inside a sample chamber drilled to a diameter of 75 µm in a rhenium gasket of 250 µm initial thickness pre-indented to 18 µm. Cerium is very likely to oxidize in open air so the sample was loaded in an argon filled glove box where both $H_2O$ and $O_2$ concentrations were maintained below 0.1 ppm. A small piece of gold (~5 µm width) was also placed near the sample for pressure calibration as shown in supplementary Fig. S14. For hydrogen loading, sample chamber was initially sealed by slightly closing the gasket and then opened in the high-pressure gas loading system in order to fill it with high purity hydrogen gas at room temperature under ~1.7 kilo-bar pressure. After hydrogen loading, Raman spectra of the $H_2$ vibron, collected at 9 GPa from the sample chamber, confirmed the presence of $H_2$ inside the sample chamber (supplementary Fig. S15). XRD patterns were recorded at beamline 13-IDD of GSECARS at the Advanced Photon Source. Angle-dispersive XRD patterns were recorded on a PILATUS CdTe 1M detector with a synchrotron radiation of incident wavelength 0.3344 Å focused to a spot size of ~3x4 µm (FWHM). A clean-up slit with an 8 µm size pin-hole was used to cut down the beam tails and collect the XRD from the smallest area possible. Pulsed laser-heating was carried out using the online infrared laser set-up with a wavelength of 1064 nm available at beamline 13-IDD[53]. Several cycles of laser heating were carried at each pressure of 33, 60, 80, 89, 98 and 100 GPa during compression and at 79 and 54 GPa during the decompression run. The use of the pulse laser heating not just helped to promote reaction between Ce and $H_2$ to form cerium hydrides but also reduced the possibility of diamond anvil failure. Pulsed-laser heating with microsecond pulse width has been utilized to reach temperatures of 1,000-2,000 K. Every laser heating shot was formed by accumulating 300k frequency modulated laser pulses of one microsecond pulse-width at a rate of 10kHz. The flat top of the laser heating spot size was around 10 µm in diameter. We strictly avoided temperatures above 2000 K to protect diamonds as hydrogen loaded DACs at high *P-T* conditions are most likely to fail[32,54]. Maintaining the sample temperature below 2000 K and relatively cold surrounding area using pulsed heating mode helped to avoid contamination and parasitic reaction with the sample chamber wall of gasket. Also, the laser heating spot on the sample was consistently maintained at substantial distance from gasket wall to avoid any contamination due to unwanted reaction (see supplementary Fig. S16). *In situ* temperature measurements were carried by fitting the slope of thermal radiation spectra to a Planck radiation function. Uncertainty in temperature measurements were less than ±100 K. Obtained raw images of XRD were integrated with DIOPTAS software[55]. Rietveld and Lebail refinements were carried out using FullPROF software.

**Calculation Details.** Evolutionary variable-composition simulation, implemented in USPEX, is used to explore the high-pressure phase diagram of the Ce-H system from ambient pressure to 250 GPa. The evolutionary algorithm USPEX[56-60] is a powerful method for finding thermodynamically stable compounds of a given system and their most stable structures. This method has been shown to be successful in predicting high-pressure structures of variety of systems which were confirmed experimentally, in specific superconducting hydrides, e.g., $UH_8$[30] and $H_3S$[14]. In this method, the first generation of structures (100 structures) and compositions are created using the random symmetric algorithm. Subsequent generations were obtained using 40% heredity, 20% transmutation, 20% softmutation, and 20% random symmetric generator. We allowed variation operators to automatically evolve in the subsequent generations. The underlying structure relaxations were carried out using the VASP package[60] in the framework of DFT and using PBE-GGA (Perdew-Burke-Ernzerhof generalized gradient approximation) [61,62]. We believe PBE is the most appropriate choice, because PBE best reproduces the experimental data (see supplementary Fig. S17 and Tables S4 and S5).The projector-augmented wave approach (PAW)[60,63] was used to describe the core electrons and their effects on valence orbitals. Valence electron configuration of $5s^2 5p^6 4f^1 5d^1 6s^2$ (i.e., with explicitly included f electrons) and $1s^1$ was used for the Ce and H atoms, respectively. A plane-wave kinetic-energy cut off of 1000 eV for hard PAW potentials and dense Monkhorst-Pack k-points grids with reciprocal space resolution of $2\pi \times 0.03$ Å$^{-1}$ were employed[64] to sample the Brillouin zone. Phonon frequencies and superconducting critical temperature were calculated using density-functional perturbation theory as implemented in the QUANTUM ESPRESSO package[65], also using the PBE-GGA functional. Ultrasoft pseudopotentials for Ce and H were used with a plane-wave basis set cut-off of 70 Ry, which gives a convergence in energy with a precision of 1 meV/atom. Phonon dispersions were also calculated under the quasi-harmonic approximation using the finite displacement method implemented in PHONOPY package[66] using forces computed with VASP. The k-space integration (electrons) was approximated by a summation over a 12 x 12 x 6 uniform grid in reciprocal space, with the smearing scheme of Methfessel-Paxton and a fictitious smearing temperature T of $k_BT = 0.05$ Ry for self-consistent cycles and relaxations; a much finer (24 x 24 x 12) grid was used for evaluating DOS and electron-phonon linewidths. Dynamical matrices and electron-phonon linewidths of *P6₃/mmc*-$CeH_9$ were calculated on a uniform 6 x 6 x 3 grid in q-space. The superconducting transition temperature $T_c$ was estimated using the Allen-Dynes modified McMillan equation[67]

$$T_c = \frac{\omega_{\log}}{1.2} \exp\left(\frac{-1.04\,(1+\lambda)}{\lambda-\mu^*(1+0.62\lambda)}\right), \quad (1)$$

where $\mu^*$ is the Coulomb pseudopotential and $\omega_{\log}$ is the logarithmic average phonon frequency. The electron-phonon coupling constant $\lambda$ and $\omega_{\log}$ were calculated as

$$\omega_{\log} = \exp\left(\frac{2}{\lambda} \int \frac{d\omega}{\omega}\, \alpha^2 F(\omega) \ln(\omega)\right), \quad (2)$$

$$\lambda = 2 \int_0^\infty \frac{\alpha^2 F(\omega)}{\omega} d\omega \quad (3)$$

**Acknowledgements**

The authors thank Suyu Fu of UT Austin for his assistance for the XRD data collection and Yuyong Xiong for hydrogen loading at Center for High Pressure Science and Technology

Advanced Research (HPSTAR). J.Z., A.O., and J.F.L acknowledge the support from DOD-ARMY grant (W911NF-16-1-0559). M.M.D.E acknowledges support from the National Science Foundation (EAR-1723160) for supporting this work. J.F.L acknowledges support from the National Science Foundation of China (NSFC) (Grant No. 11872077). N.P.S., Y.Z. and J.F.L. acknowledges support from HPSTAR. Y.Z. acknowledges support from "the Fundamental Research Funds for the central universities" in China and from the NSFC (Grant No. 41804082). High-pressure XRD experiments were conducted at GeoSoilEnviroCARS of APS, ANL. GeoSoilEnviroCARS operations are supported by the National Science Foundation-Earth Sciences (EAR-1128799) and the Department of Energy, Geosciences (DE-FG02-94ER14466). This research used resources of the Advanced Photon Source, a U.S. Department of Energy (DOE) Office of Science User Facility operated for the DOE Office of Science by Argonne National Laboratory under Contract No. DE-AC02-06CH11357. Calculations were mainly performed on the cluster (QSH) in our laboratory at Stony Brook University.

## Author contributions

J.F.L. and N.P.S. conceived the project. J.F.L. supervised the project, and J.F.L. and N.P.S. coordinated the project. N.P.S. planned the experiment and prepared the diamond anvil cell samples. Y.Z., J.F.L., E.G. and V.B.P. carried out the XRD measurements with the laser heating at high pressures. N.P.S. analysed and modelled the data, N.P.S. and J.F.L. interpreted the initial data. M.M.D.E., I.A.K. and A.R.O. carried out the theoretical calculations. All the authors discussed the results. N.P.S. and M. M. D. E. wrote the draft. A.R.O., and J.F.L. participated in writing and revising the manuscript, and J.Z. and J.F.L. helped with the draft abstract. All the authors read and commented on the manuscript.

## References

1 Wigner, E. & Huntington, H. B. On the Possibility of a Metallic Modification of Hydrogen. *The Journal of Chemical Physics* **3**, 764-770, doi:10.1063/1.1749590 (1935).

2 Ashcroft, N. W. Metallic Hydrogen: A High-Temperature Superconductor? *Physical Review Letters* **21**, 1748-1749 (1968).

3 Loubeyre, P., Occelli, F. & LeToullec, R. Optical studies of solid hydrogen to 320 GPa and evidence for black hydrogen. *Nature* **416**, 613, doi:10.1038/416613a (2002).

4 Eremets, M. I. & Troyan, I. A. Conductive dense hydrogen. *Nature Materials* **10**, 927, doi:10.1038/nmat3175 (2011).

5 Dalladay-Simpson, P., Howie, R. T. & Gregoryanz, E. Evidence for a new phase of dense hydrogen above 325 gigapascals. *Nature* **529**, 63, doi:10.1038/nature16164 (2016).

6 Dias, R. P. & Silvera, I. F. Observation of the Wigner-Huntington transition to metallic hydrogen. *Science* (2017).

7 Borinaga, M., Errea, I., Calandra, M., Mauri, F. & Bergara, A. Anharmonic effects in atomic hydrogen: Superconductivity and lattice dynamical stability. *Physical Review B* **93**, 174308 (2016).

8 Azadi, S., Monserrat, B., Foulkes, W. M. C. & Needs, R. J. Dissociation of High-Pressure Solid Molecular Hydrogen: A Quantum Monte Carlo and Anharmonic Vibrational Study. *Physical Review Letters* **112**, 165501 (2014).

9 McMinis, J., Clay, R. C., Lee, D. & Morales, M. A. Molecular to Atomic Phase Transition in Hydrogen under High Pressure. *Physical Review Letters* **114**, 105305 (2015).

10 Gor'kov, L. P. & Kresin, V. Z. Colloquium: High pressure and road to room temperature superconductivity. *Reviews of Modern Physics* **90**, 011001 (2018).

11 Ashcroft, N. W. Hydrogen Dominant Metallic Alloys: High Temperature Superconductors? *Physical Review Letters* **92**, 187002 (2004).

12 Struzhkin, V. V. Superconductivity in compressed hydrogen-rich materials: Pressing on hydrogen. *Physica C: Superconductivity and its Applications* **514**, 77-85, doi:https://doi.org/10.1016/j.physc.2015.02.017 (2015).

13 McMahon, J. M. & Ceperley, D. M. High-temperature superconductivity in atomic metallic hydrogen. *Physical Review B* **84**, 144515 (2011).

14 Duan, D. *et al.* Pressure-induced metallization of dense (H2S)2H2 with high-Tc superconductivity. *Scientific Reports* **4**, 6968, doi:10.1038/srep06968 (2014).

15 Drozdov, A. P., Eremets, M. I., Troyan, I. A., Ksenofontov, V. & Shylin, S. I. Conventional superconductivity at 203 kelvin at high pressures in the sulfur hydride system. *Nature* **525**, 73, doi:10.1038/nature14964 (2015).

16 Shaw, B. L. in *Inorganic Hydrides* 1-3 (Pergamon, 1967).

17 Blackledge, J. P. in *Metal Hydrides* 1-20 (Academic Press, 1968).

18 Zurek, E. Hydrides of the Alkali Metals and Alkaline Earth Metals Under Pressure. *Comments on Inorganic Chemistry* **37**, 78-98, doi:10.1080/02603594.2016.1196679 (2017).

19 Zeng, Q., Yu, S., Li, D., Oganov, A. R. & Frapper, G. Emergence of novel hydrogen chlorides under high pressure. *Physical Chemistry Chemical Physics* **19**, 8236-8242, doi:10.1039/C6CP08708F (2017).

20 Mahdi Davari Esfahani, M. *et al.* Superconductivity of novel tin hydrides (SnnHm) under pressure. *Scientific Reports* **6**, 22873, doi:10.1038/srep22873 (2016).

21 Peng, F. *et al.* Hydrogen Clathrate Structures in Rare Earth Hydrides at High Pressures: Possible Route to Room-Temperature Superconductivity. *Physical Review Letters* **119**, 107001 (2017).

22 Wang, H., Tse, J. S., Tanaka, K., Iitaka, T. & Ma, Y. Superconductive sodalite-like clathrate calcium hydride at high pressures. *Proceedings of the National Academy of Sciences* **109**, 6463-6466, doi:10.1073/pnas.1118168109 (2012).

23 Lonie, D. C., Hooper, J., Altintas, B. & Zurek, E. Metallization of magnesium polyhydrides under pressure. *Physical Review B* **87**, 054107 (2013).

24 Feng, X., Zhang, J., Gao, G., Liu, H. & Wang, H. Compressed sodalite-like MgH6 as a potential high-temperature superconductor. *RSC Advances* **5**, 59292-59296, doi:10.1039/C5RA11459D (2015).

25 Li, Y. *et al.* Pressure-stabilized superconductive yttrium hydrides. *Scientific Reports* **5**, 9948, doi:10.1038/srep09948 (2015).

26 Liu, H., Naumov, I. I., Hoffmann, R., Ashcroft, N. W. & Hemley, R. J. Potential high Tc superconducting lanthanum and yttrium hydrides at high pressure. *Proceedings of the National Academy of Sciences* **114**, 6990-6995, doi:10.1073/pnas.1704505114 (2017).

27 Semenok, D. V., Kvashnin, A. G., Kruglov, I. A. & Oganov, A. R. Actinium Hydrides AcH10, AcH12, and AcH16 as High-Temperature Conventional Superconductors. *The Journal of Physical Chemistry Letters*, 1920-1926, doi:10.1021/acs.jpclett.8b00615 (2018).

28 Pépin, C. M., Geneste, G., Dewaele, A., Mezouar, M. & Loubeyre, P. Synthesis of FeH5: A layered structure with atomic hydrogen slabs. *Science* **357**, 382-385, doi:10.1126/science.aan0961 (2017).
29 Geballe, Z. M. *et al.* Synthesis and Stability of Lanthanum Superhydrides. *Angewandte Chemie International Edition* **57**, 688-692, doi:10.1002/anie.201709970 (2018).
30 Kruglov, I. A. *et al.* Uranium polyhydrides at moderate pressures: prediction, synthesis, and expected superconductivity. *arXiv:1708.05251* (2017).
31 Pépin, C., Loubeyre, P., Occelli, F. & Dumas, P. Synthesis of lithium polyhydrides above 130 GPa at 300 K. *Proceedings of the National Academy of Sciences* **112**, 7673-7676, doi:10.1073/pnas.1507508112 (2015).
32 Struzhkin, V. V. *et al.* Synthesis of sodium polyhydrides at high pressures. *Nature Communications* **7**, 12267, doi:10.1038/ncomms12267 (2016).
33 Somayazulu, M. *et al.* Pressure-induced bonding and compound formation in xenon–hydrogen solids. *Nature Chemistry* **2**, 50, doi:10.1038/nchem.445 (2009).
34 Binns, J. *et al.* Formation of ${\mathrm{H}}_{2}$-rich iodine-hydrogen compounds at high pressure. *Physical Review B* **97**, 024111 (2018).
35 Somayazulu, M. *et al.* Evidence for superconductivity above 260 K in lanthanum superhydride at megabar pressures. *arXIv* **arXiv:1808.07695** (2018).
36 Drozdov, A. P. *et al.* Superconductivity at 215 K in lanthanum hydride at high pressures. *arXiv* **arXiv:1808.07039** (2018).
37 Crabtree, G. W., Dresselhaus, M. S. & Buchanan, M. V. The Hydrogen Economy. *Physics Today* **57**, 39-44, doi:10.1063/1.1878333 (2004).
38 Halevy, I., Salhov, S., Zalkind, S., Brill, M. & Yaar, I. High pressure study of β-UH3 crystallographic and electronic structure. *Journal of Alloys and Compounds* **370**, 59-64, doi:https://doi.org/10.1016/j.jallcom.2003.09.124 (2004).
39 Ce, M. *et al.* Structure phase transformation and equation of state of cerium metal under pressures up to 51 GPa. *Chinese Physics B* **25**, 046401 (2016).
40 Lipp, M. J. *et al.* Comparison of the high-pressure behavior of the cerium oxides $\mathrm{C}{\mathrm{e}}_{2}{\mathrm{O}}_{3}$ and $\mathrm{Ce}{\mathrm{O}}_{2}$. *Physical Review B* **93**, 064106 (2016).
41 Jacobsen, M. K., Velisavljevic, N., Dattelbaum, D. M., Chellappa, R. S. & Park, C. High pressure and temperature equation of state and spectroscopic study of CeO 2. *Journal of Physics: Condensed Matter* **28**, 155401 (2016).
42 Davari Esfahani, M. M., Oganov, A. R., Niu, H. & Zhang, J. Superconductivity and unexpected chemistry of germanium hydrides under pressure. *Physical Review B* **95**, 134506, doi:10.1103/PhysRevB.95.134506 (2017).
43 Loubeyre, P. *et al.* X-ray diffraction and equation of state of hydrogen at megabar pressures. *Nature* **383**, 702, doi:10.1038/383702a0 (1996).
44 Pépin, C. M., Dewaele, A., Geneste, G., Loubeyre, P. & Mezouar, M. New Iron Hydrides under High Pressure. *Physical Review Letters* **113**, 265504 (2014).
45 Goncharenko, I. *et al.* Pressure-Induced Hydrogen-Dominant Metallic State in Aluminum Hydride. *Physical Review Letters* **100**, 045504 (2008).
46 McMahon, J. M. & Ceperley, D. M. Ground-State Structures of Atomic Metallic Hydrogen. *Physical Review Letters* **106**, 165302 (2011).
47 Tanaka, K., Tse, J. S. & Liu, H. Electron-phonon coupling mechanisms for hydrogen-rich metals at high pressure. *Physical Review B* **96**, 100502, doi:10.1103/PhysRevB.96.100502 (2017).
48 Vohra, Y. K., Beaver, S. L., Akella, J., Ruddle, C. A. & Weir, S. T. Ultrapressure equation of state of cerium metal to 208 GPa. *Journal of Applied Physics* **85**, 2451-2453, doi:10.1063/1.369566 (1999).

49 Carbotte, J. P., Nicol, E. J. & Timusk, T. Detecting Superconductivity in the High Pressure Hydrides and Metallic Hydrogen from Optical Properties. *Physical Review Letters* **121**, 047002, doi:10.1103/PhysRevLett.121.047002 (2018).
50 Zhu, Q., Oganov, A. R., Lyakhov, A. O. & Yu, X. Generalized evolutionary metadynamics for sampling the energy landscapes and its applications. *Physical Review B* **92**, 024106 (2015).
51 Errea, I. *et al.* High-Pressure Hydrogen Sulfide from First Principles: A Strongly Anharmonic Phonon-Mediated Superconductor. *Physical Review Letters* **114**, 157004, doi:10.1103/PhysRevLett.114.157004 (2015).
52 Errea, I. *et al.* Quantum hydrogen-bond symmetrization in the superconducting hydrogen sulfide system. *Nature* **532**, 81, doi:10.1038/nature17175 (2016).
53 Prakapenka, V. B. *et al.* Advanced flat top laser heating system for high pressure research at GSECARS: application to the melting behavior of germanium. *High Pressure Research* **28**, 225-235, doi:10.1080/08957950802050718 (2008).
54 Goncharov, A. F. *et al.* X-ray diffraction in the pulsed laser heated diamond anvil cell. *Review of Scientific Instruments* **81**, 113902, doi:10.1063/1.3499358 (2010).
55 Prescher, C. & Prakapenka, V. B. DIOPTAS: a program for reduction of two-dimensional X-ray diffraction data and data exploration. *High Pressure Research* **35**, 223-230, doi:10.1080/08957959.2015.1059835 (2015).
56 Glass, C. W., Oganov, A. R. & Hansen, N. USPEX—Evolutionary crystal structure prediction. *Computer Physics Communications* **175**, 713-720, doi:https://doi.org/10.1016/j.cpc.2006.07.020 (2006).
57 Oganov, A. R. & Glass, C. W. Crystal structure prediction using ab initio evolutionary techniques: Principles and applications. *The Journal of Chemical Physics* **124**, 244704, doi:10.1063/1.2210932 (2006).
58 Lyakhov, A. O., Oganov, A. R. & Valle, M. How to predict very large and complex crystal structures. *Computer Physics Communications* **181**, 1623-1632, doi:https://doi.org/10.1016/j.cpc.2010.06.007 (2010).
59 Oganov, A. R., Lyakhov, A. O. & Valle, M. How Evolutionary Crystal Structure Prediction Works—and Why. *Accounts of Chemical Research* **44**, 227-237, doi:10.1021/ar1001318 (2011).
60 Kresse, G. & Joubert, D. From ultrasoft pseudopotentials to the projector augmented-wave method. *Physical Review B* **59**, 1758-1775 (1999).
61 Perdew, J. P., Burke, K. & Ernzerhof, M. Generalized Gradient Approximation Made Simple. *Physical Review Letters* **77**, 3865-3868 (1996).
62 Perdew, J. P., Burke, K. & Ernzerhof, M. Generalized Gradient Approximation Made Simple [Phys. Rev. Lett. 77, 3865 (1996)]. *Physical Review Letters* **78**, 1396-1396 (1997).
63 Blöchl, P. E. Projector augmented-wave method. *Physical Review B* **50**, 17953-17979 (1994).
64 Monkhorst, H. J. & Pack, J. D. Special points for Brillouin-zone integrations. *Physical Review B* **13**, 5188-5192 (1976).
65 Paolo, G. *et al.* QUANTUM ESPRESSO: a modular and open-source software project for quantum simulations of materials. *Journal of Physics: Condensed Matter* **21**, 395502 (2009).
66 Togo, A. & Tanaka, I. First principles phonon calculations in materials science. *Scripta Materialia* **108**, 1-5, doi:https://doi.org/10.1016/j.scriptamat.2015.07.021 (2015).
67 Allen, P. B. & Dynes, R. C. Transition temperature of strong-coupled superconductors reanalyzed. *Physical Review B* **12**, 905-922 (1975).

# Supplementary Information for

## Synthesis of clathrate cerium superhydride $CeH_9$ below 100 GPa with atomic hydrogen sublattice

Nilesh P. Salke[1], M. Mahdi Davari Esfahani[2], Youjun Zhang[3,1], Ivan A. Kruglov[4,5], Jianshi Zhou[6], Yaguo Wang[6], Eran Greenberg[7], Vitali B. Prakapenka[7], Artem R. Oganov[8,4,9,*], Jung-Fu Lin[10,*]

[1]Center for High Pressure Science & Technology Advanced Research (HPSTAR), Shanghai, 201203, China
[2]Department of Geosciences, Center for Materials by Design, and Institute for Advanced Computational Science, State University of New York, Stony Brook, New York 11794-2100, USA
[3]Institute of Atomic and Molecular Physics, Sichuan University, Chengdu 610065, China
[4]Department of Problems of Physics and Energetics, Moscow Institute of Physics and Technology, 9 Institutskiy Lane, Dolgoprudny City, Moscow Region 141700, Russia
[5]Dukhov Research Institute of Automatics (VNIIA), Moscow 127055, Russia
[6]Department of Mechanical Engineering, The University of Texas at Austin, Austin, Texas 78712, USA
[7]Center for Advanced Radiation Sources, University of Chicago, Illinois, 60637, USA
[8]Skolkovo Institute of Science and Technology, Skolkovo Innovation Center, 3 Nobel Street, Moscow 143026, Russia
[9]International Center for Materials Design, Northwestern Polytechnical University, Xi'an 710072, China
[10]Department of Geological Sciences, The University of Texas at Austin, Austin, Texas 78712, USA

XRD pattern of the sample initially loaded at 9 GPa could be indexed with $Fm\bar{3}m$ structure. From the literature, it is well known that at ambient *P-T* conditions $CeH_X$ system can stabilize in various stoichiometries ranging from X = 2 to 3[1]. Specifically, two extreme compositions $CeH_2$ and $CeH_3$ stabilize in $Fm\bar{3}m$ structure whereas $CeH_{2.5}$ ($Ce_2H_5$) forms tetragonal structure with $I4_1md$ space group[1]. Besides this, $CeO_2$, which could occur due to starting sample being contaminated and oxidized, also has fluorite type $Fm\bar{3}m$ structure[2]. It was difficult to confirm the exact stoichiometry and composition of observed $Fm\bar{3}m$ phase at 9 GPa. It was also important to confirm stoichiometry and composition of the sample to rule out any contamination. Reported values of lattice parameter at ambient pressure for $Fm\bar{3}m$ structured $CeH_2$ and $CeH_3$ were 5.580 and 5.534 Å respectively[3]; corresponding volume per formula unit (V/f.u.) is 43.435 and 42.37 $Å^3$ respectively[3]. We fitted the experimental V/f.u. (assuming Z=4) of observed $Fm\bar{3}m$ phase using third order Birch-Murnaghan EOS (see SI Figure S1d). Zero pressure volume obtained after fitting is $V_0$ = 44(1) $Å^3$/f.u., which is closer to the reported $V_0$/f.u. of $CeH_2$, which rules out the $Fm\bar{3}m$-$CeH_3$ phase. In Fig. S1d, we also compared the experimental EOS with reported EOS of $CeO_2$[2]. From Fig. S1d, it clearly indicates that experimental EOS of $Fm\bar{3}m$ phase is not consistent with EOS of $Fm\bar{3}m$-$CeO_2$ which firmly rules out the presence of $Fm\bar{3}m$-$CeO_2$ in the sample. Hence, we concluded that $Fm\bar{3}m$ structure observed in the beginning at 9 GPa was of $CeH_2$ phase which persisted up to 33 GPa until we carried laser heating on the sample.

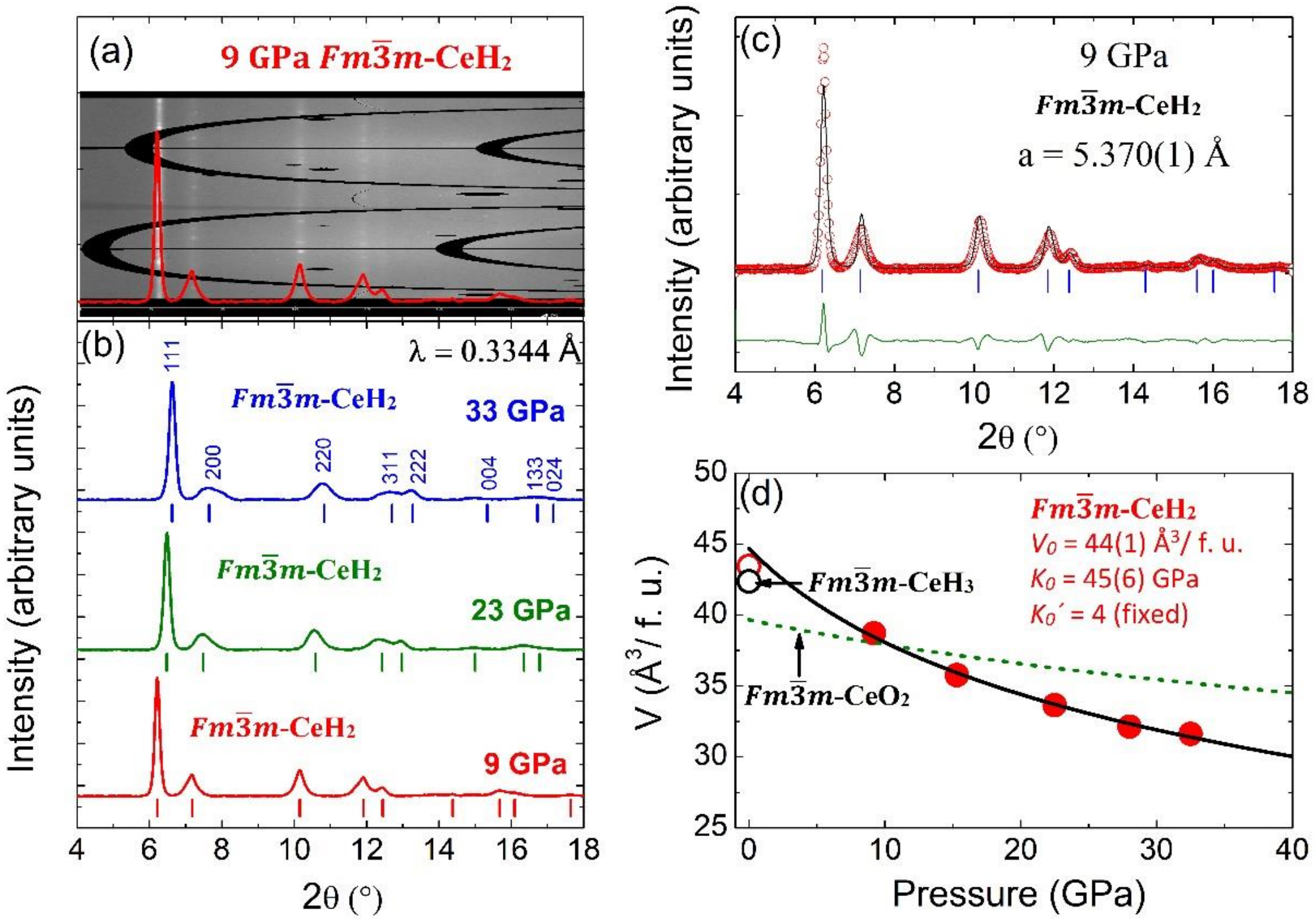

**Supplementary Figure S1: Representative x-ray diffraction patterns at high pressures and equation of state of $Fm\bar{3}m$-$CeH_2$.** (a) XRD image at 9 GPa representing polycrystalline nature of $Fm\bar{3}m$-$CeH_2$ phase. (b) X-ray diffraction patterns of $Fm\bar{3}m$-$CeH_2$ from 9 to 33 GPa. Red, green and blue vertical ticks correspond to peak positions of $Fm\bar{3}m$-$CeH_2$ phase at 9, 23 and 33 GPa respectively. (c) Lebail refinement plot for $Fm\bar{3}m$-$CeH_2$ at 9 GPa, $\chi^2 = 5.41$ (d) Volume per formula unit of $Fm\bar{3}m$-$CeH_2$ as a function of pressure is fitted using third-order Birch-Murnaghan equation of state (EOS). Red solid circles represent experimental unit cell volume per formula unit data. Black line is EOS fit to experimental data. Red and black open circle represents reported volume per formula unit of $Fm\bar{3}m$ structured $CeH_2$ and $CeH_3$ at ambient pressure respectively.[3] Green dotted line represents reported EOS for $Fm\bar{3}m$-$CeO_2$.[2]

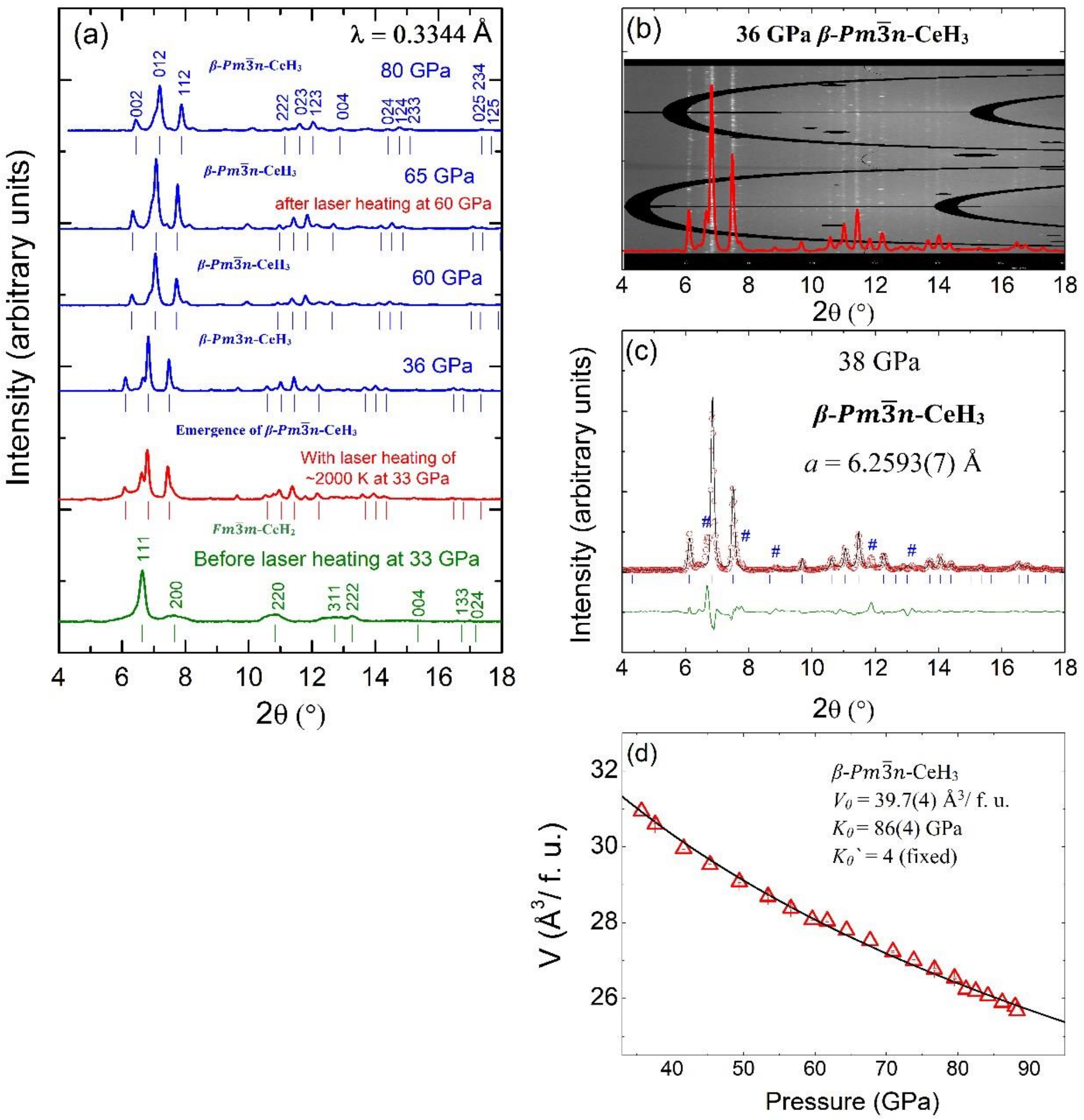

**Supplementary Figure S2: Representative x-ray diffraction patterns denoting evolution of $\beta$-$Pm\overline{3}n$-$CeH_3$ under pressure with laser heating and equation of state fitting.** (a) X-ray diffraction patterns at various pressures representing the formation of $\beta$-$Pm\overline{3}n$-$CeH_3$ up to 80 GPa. $\beta$-$Pm\overline{3}n$-$CeH_3$ phase forms after laser heating at 33 GPa and 2100 K. Vertical green ticks indicate indexing of peak position for $Fm\overline{3}m$-$CeH_2$; vertical red and blue ticks indicate indexing of peak position for $\beta$-$Pm\overline{3}n$-$CeH_3$ at respective pressure. (b) XRD image at 36 GPa representing polycrystalline nature of $\beta$-$Pm\overline{3}n$-$CeH_3$. (c) Lebail refinement plot for $\beta$-$Pm\overline{3}n$-$CeH_3$ at 38 GPa, $\chi^2 = 10.6$, # represents unidentified peaks (d) Third-order Birch-Murnaghan equation of state fitting for $\beta$-$Pm\overline{3}n$-$CeH_3$. Red open triangles represent experimental data for $\beta$-$Pm\overline{3}n$-$CeH_3$, black solid line represent EOS fit. Errors are also plotted in (d) but are mostly too small to be seen.

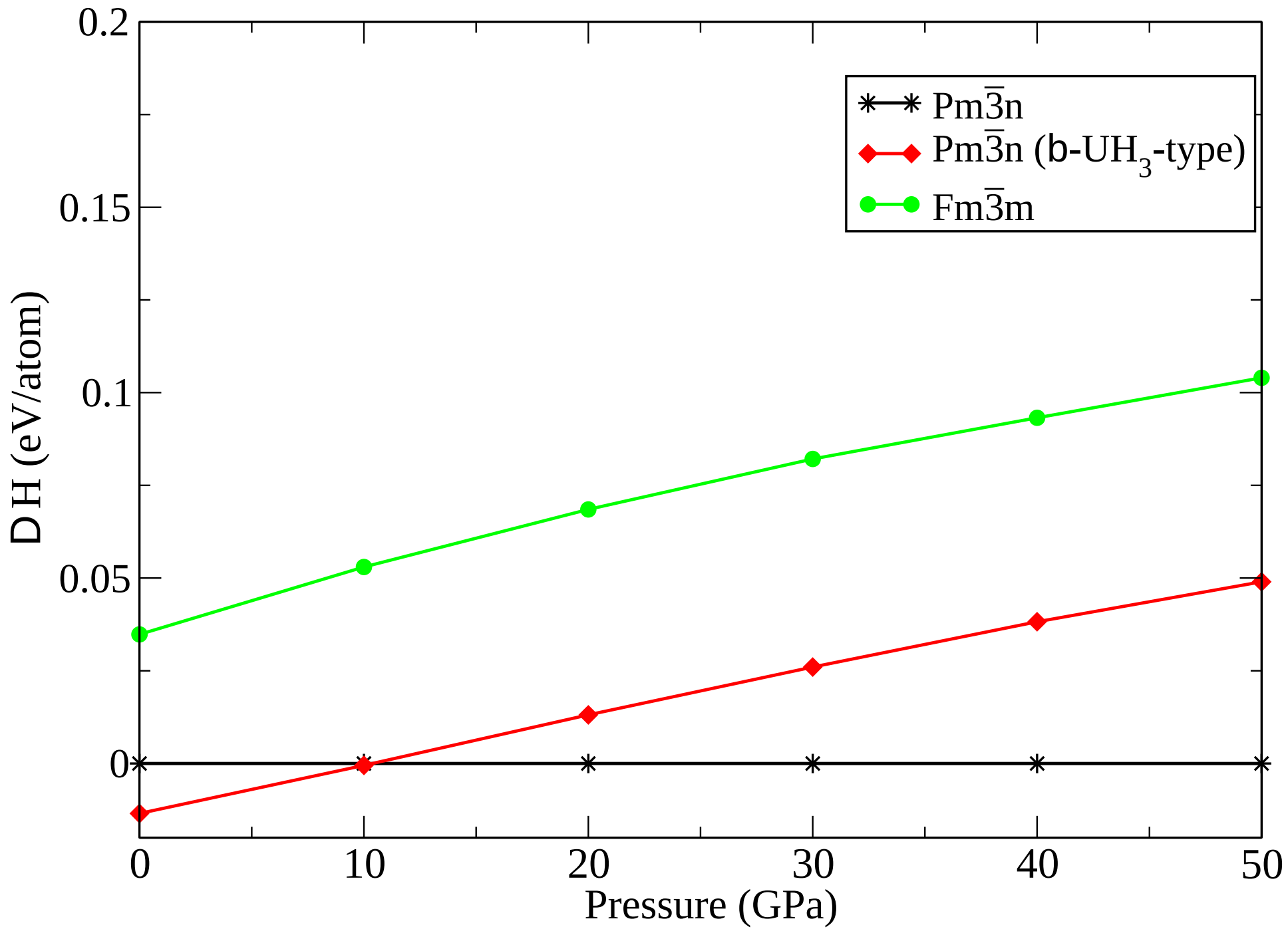

**Supplementary Figure S3: Enthalpy comparison for various $CeH_3$ phases as a function of pressure.** Enthalpy per atom relative to the $Pm\bar{3}n$ structure as a function of pressure for the best phases with the $CeH_3$ stoichiometry. Experimentally known ambient pressure phase $Fm\bar{3}m$-$CeH_3$ is added for comparison. Red, black and green symbols-line represents $\beta$-$Pm\bar{3}n$, $Pm\bar{3}n$ and $Fm\bar{3}m$ phases of $CeH_3$ respectively.

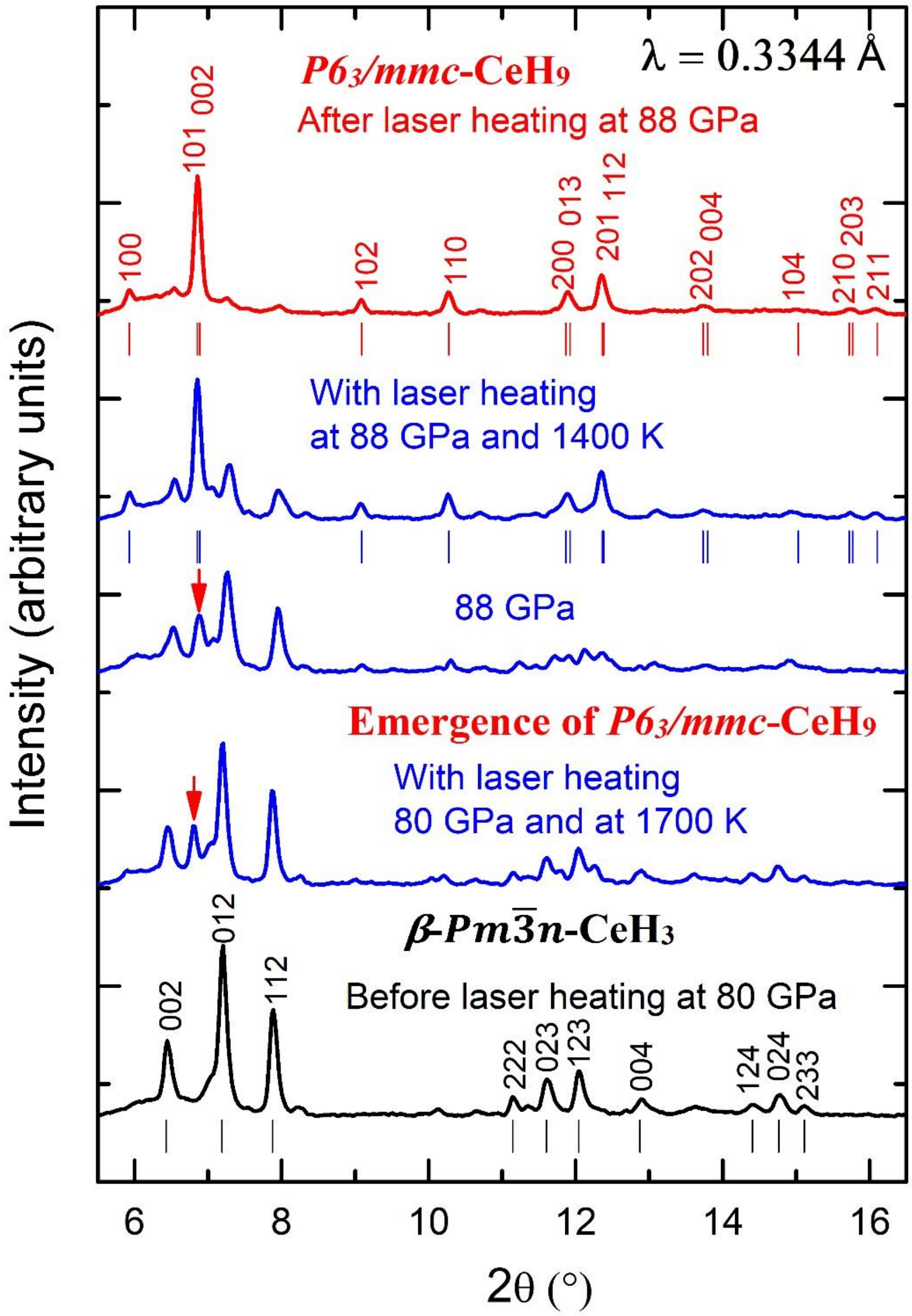

**Supplementary Figure S4: X-ray diffraction patterns representing formation of superhydride *P6₃/mmc*-$CeH_9$.** X-ray diffraction patterns at pressures above 80 GPa representing the formation of $CeH_9$ with laser heating. New peaks, which emerged after laser heating to 1700 K at 80 GPa, are shown with a red arrow corresponding to (101) and (002) of the $P6_3/mmc$ phase. Vertical black ticks indicate indexing for $\beta$-$Pm\overline{3}n$-$CeH_3$ at 80 GPa. Vertical blue and red ticks indicate indexing of the $P6_3/mmc$-$CeH_9$ phase at the respective pressures.

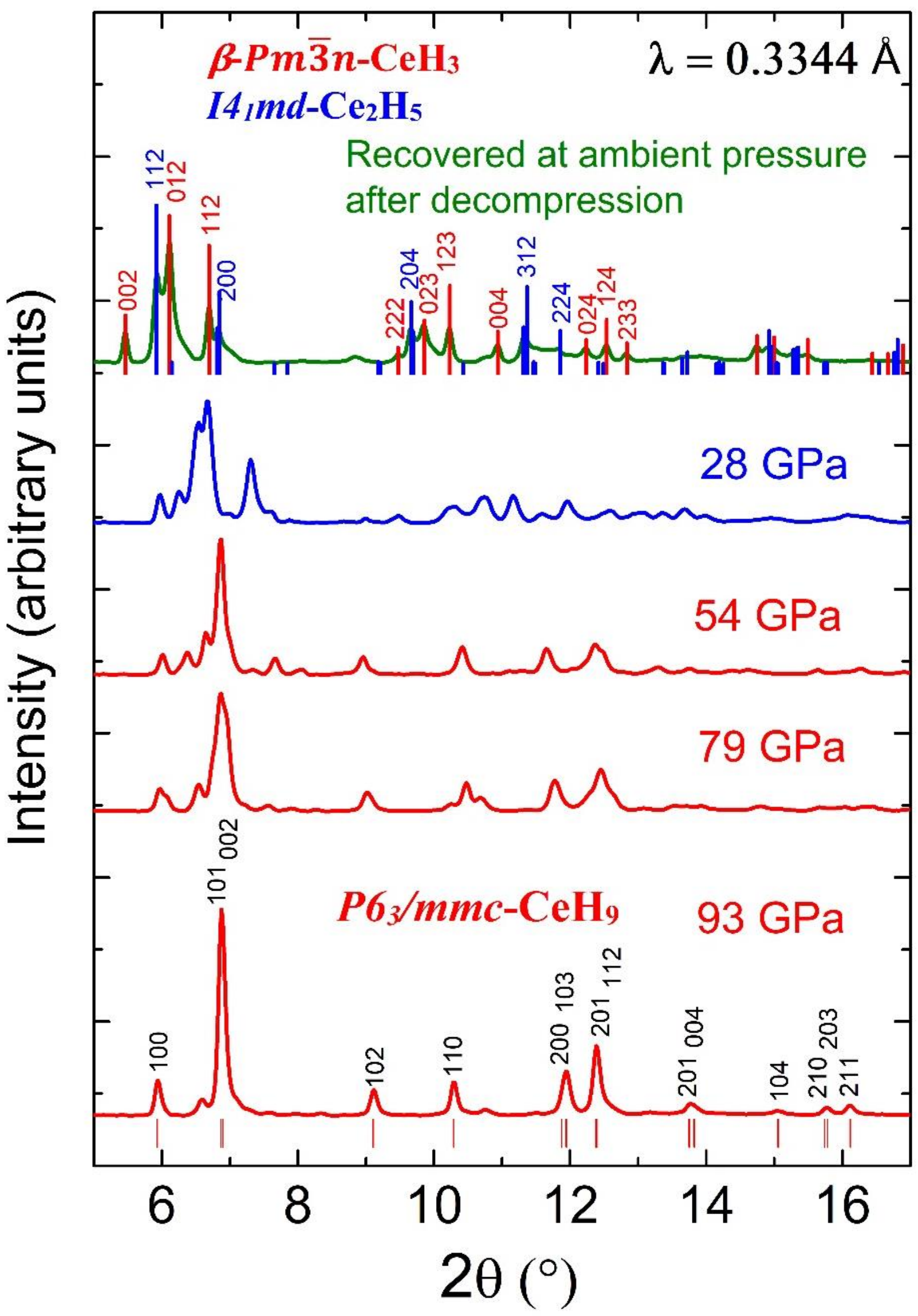

**Supplementary Figure S5: Representative x-ray diffraction patterns obtained during the decompression cycle.** XRD patterns were recorded upon the decompression cycle down to ambient pressure. Ambient pressure diffraction patterns of the recovered sample were recorded from the sample chamber after complete decompression. $Pm\bar{3}n$-$CeH_3$ and $I4_1md$-$Ce_2H_5$ phases were recovered at ambient conditions. Red vertical ticks for XRD at 93 GPa represent peak positions for $P6_3/mmc$-$CeH_9$. Red and blue vertical lines for XRD of the recovered sample at ambient pressure indicate peak positions with relative intensity for $Pm\bar{3}n$-$CeH_3$ and $I4_1md$-$Ce_2H_5$, respectively.

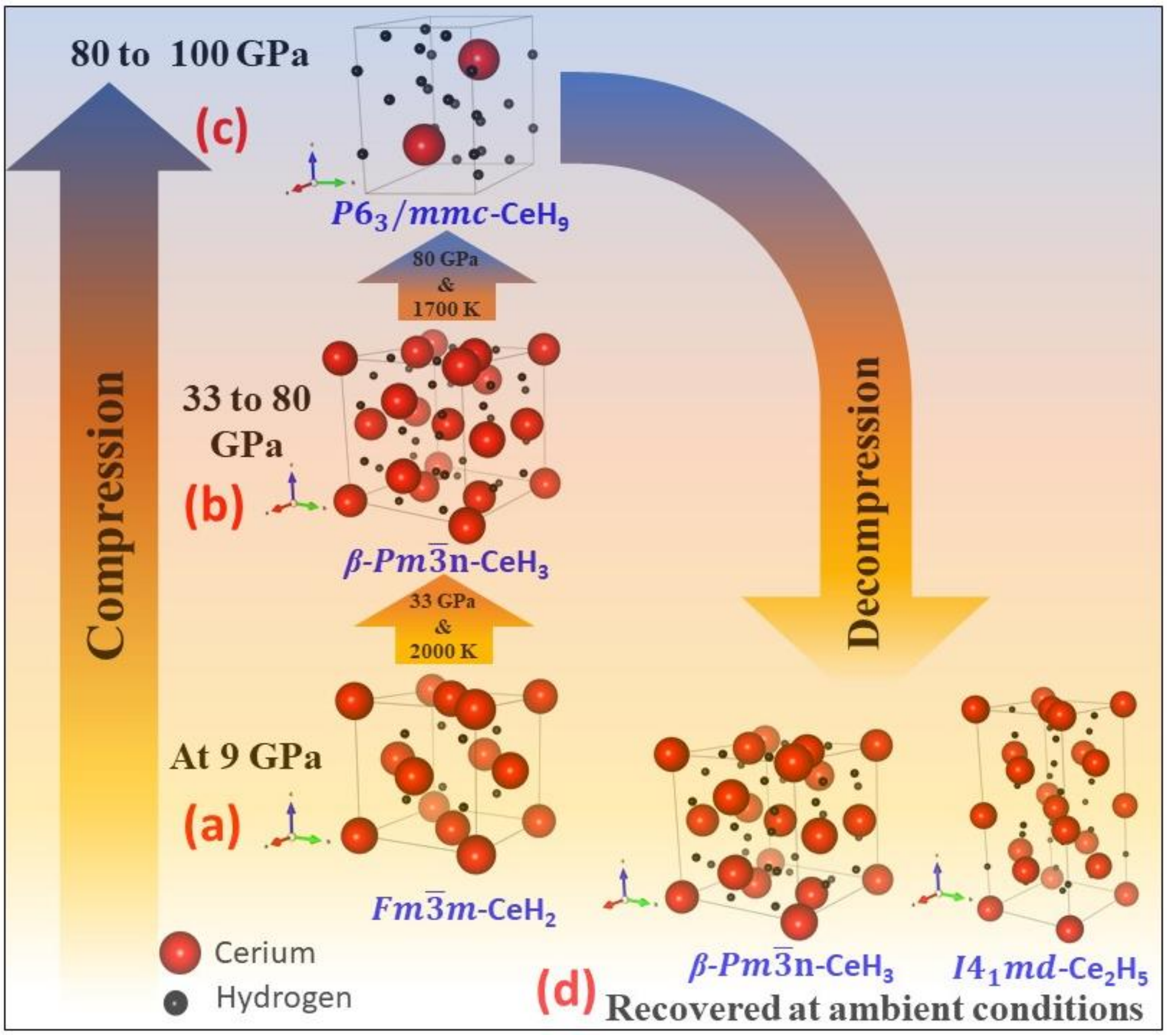

**Supplementary Figure S6: Pressure temperature path for the synthesis and stability of various Ce-H phases.** (a) Starting at 9 GPa, cerium reacts with hydrogen to form $Fm\overline{3}m$-$CeH_2$, which remained stable up to 33 GPa. (b) At 33 GPa with laser heating, $Fm\overline{3}m$-$CeH_2$ in $H_2$ medium reacted to form $\beta$-$Pm\overline{3}n$-$CeH_3$. $\beta$-$Pm\overline{3}n$-$CeH_3$ remained stable up to 80 GPa (c) Laser heating of $\beta$-$Pm\overline{3}n$-$CeH_3$ in $H_2$ medium above 80 GPa resulted in the discovery of the $P6_3/mmc$-$CeH_9$ superhydride. The superhydride phase was found to be stable up to the maximum pressure reached in our studies i.e. 100 GPa. (d) After complete decompression, $\beta$-$Pm\overline{3}n$-$CeH_3$ and $I4_1md$-$Ce_2H_5$ were recovered at ambient conditions.

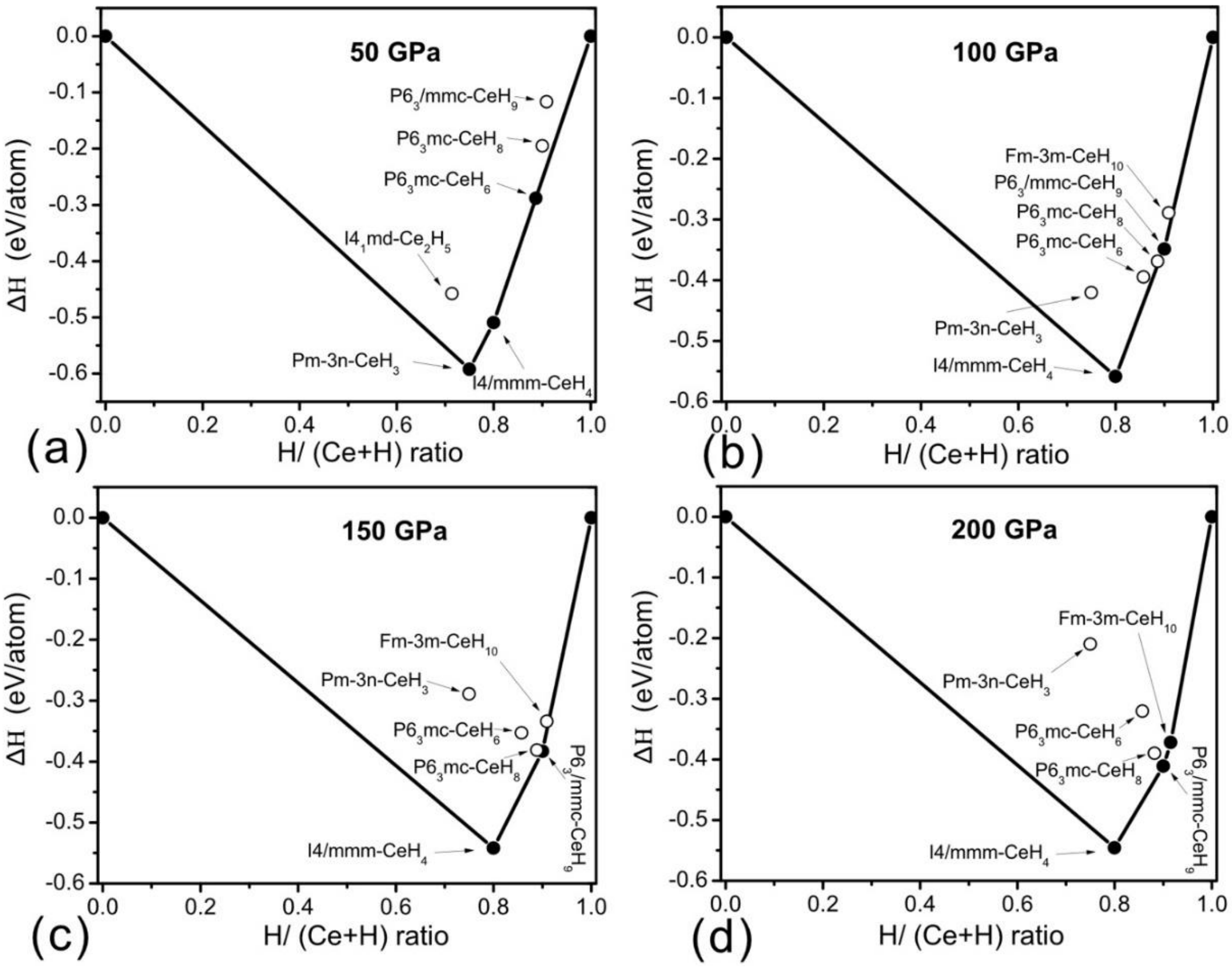

**Supplementary Figure S7: Convex Hull diagram of Ce-H system.** Predicted formation enthalpy of $Ce_{1-x}$ $H_x$ as a function of H concentration at selected pressures. Open circles above the convex hull show unstable compounds with respect to decomposition into the two adjacent phases on the convex hull, while solid circles show thermodynamically stable compounds.

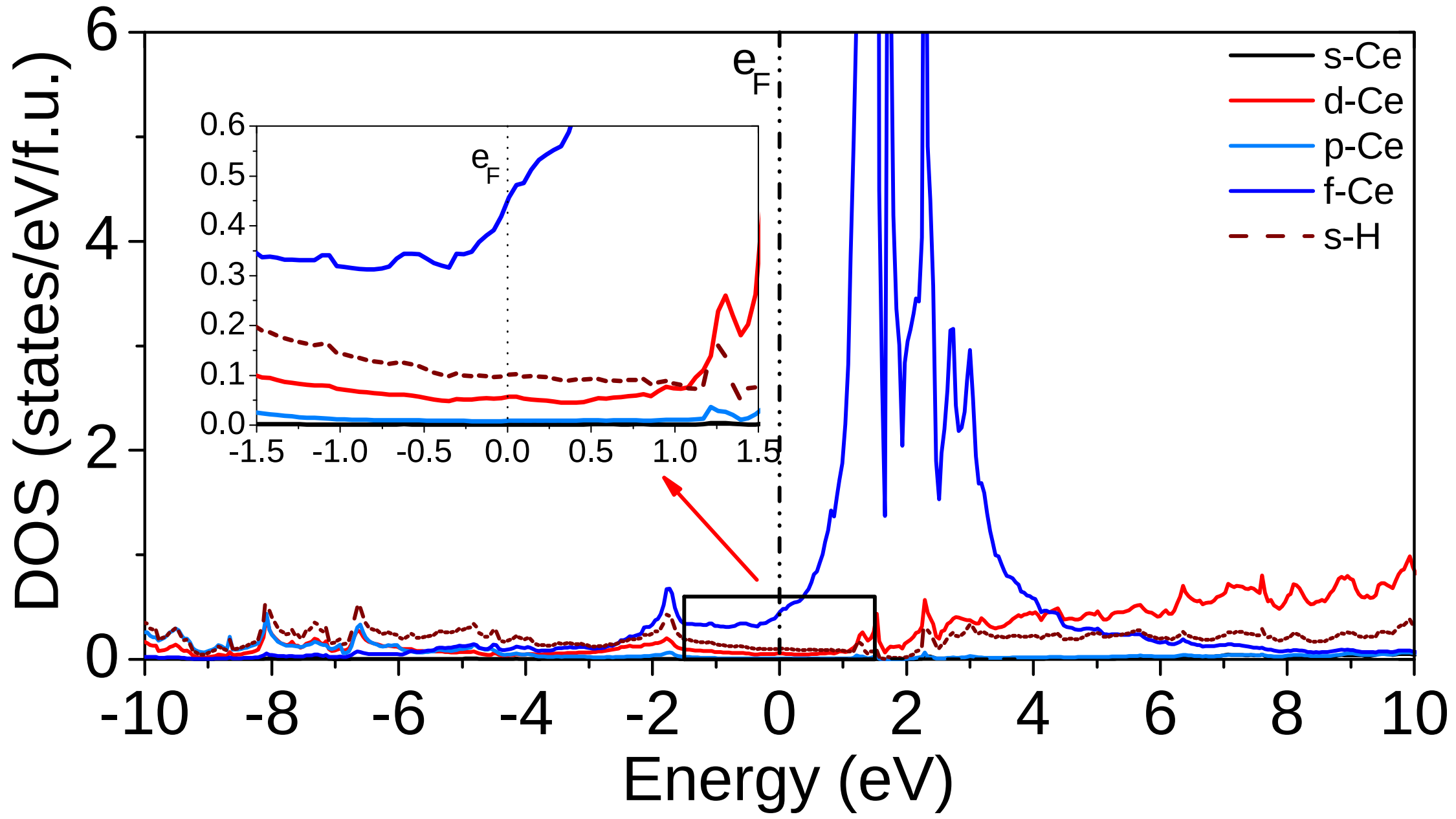

**Figure S8: Electron density of states (DOS) for $CeH_9$ at 150 GPa.** Electron DOS at Fermi level is largely dominated by Ce-4f and H orbitals.

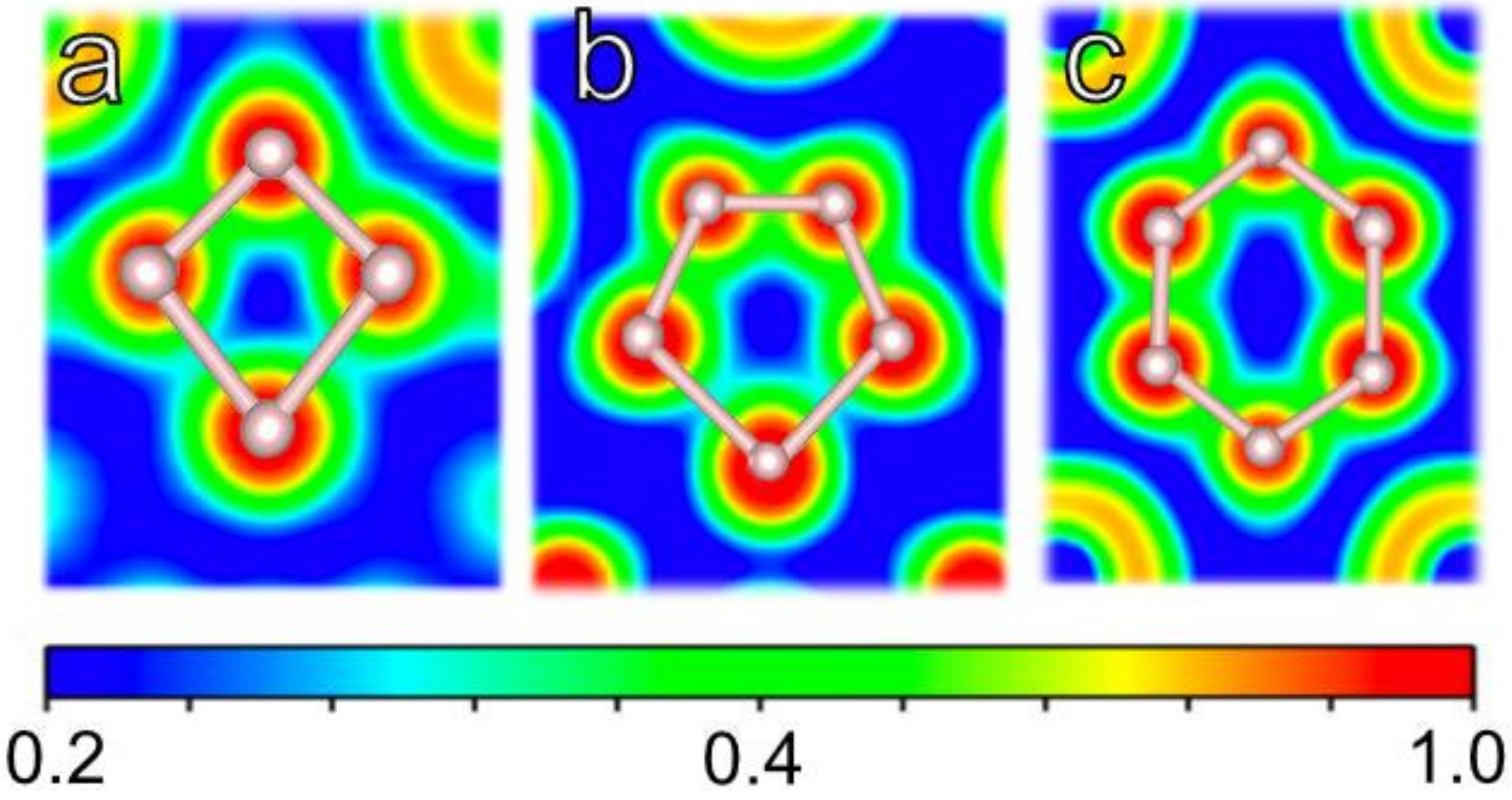

**Figure S9: Electron localization function (ELF) of $CeH_9$ at 150 GPa.** Figure (a), (b) and (c) represents ELF plots in (1.15 0 1), (1 1 0) and (-1 0 2.55) sections for $CeH_9$ at 150 GPa. Hydrogen atoms and $H_4$, $H_5$ and $H_6$ rings are shown in the figures. For better clarity, we set the minimum of the ELF to 0.2 in the plots.

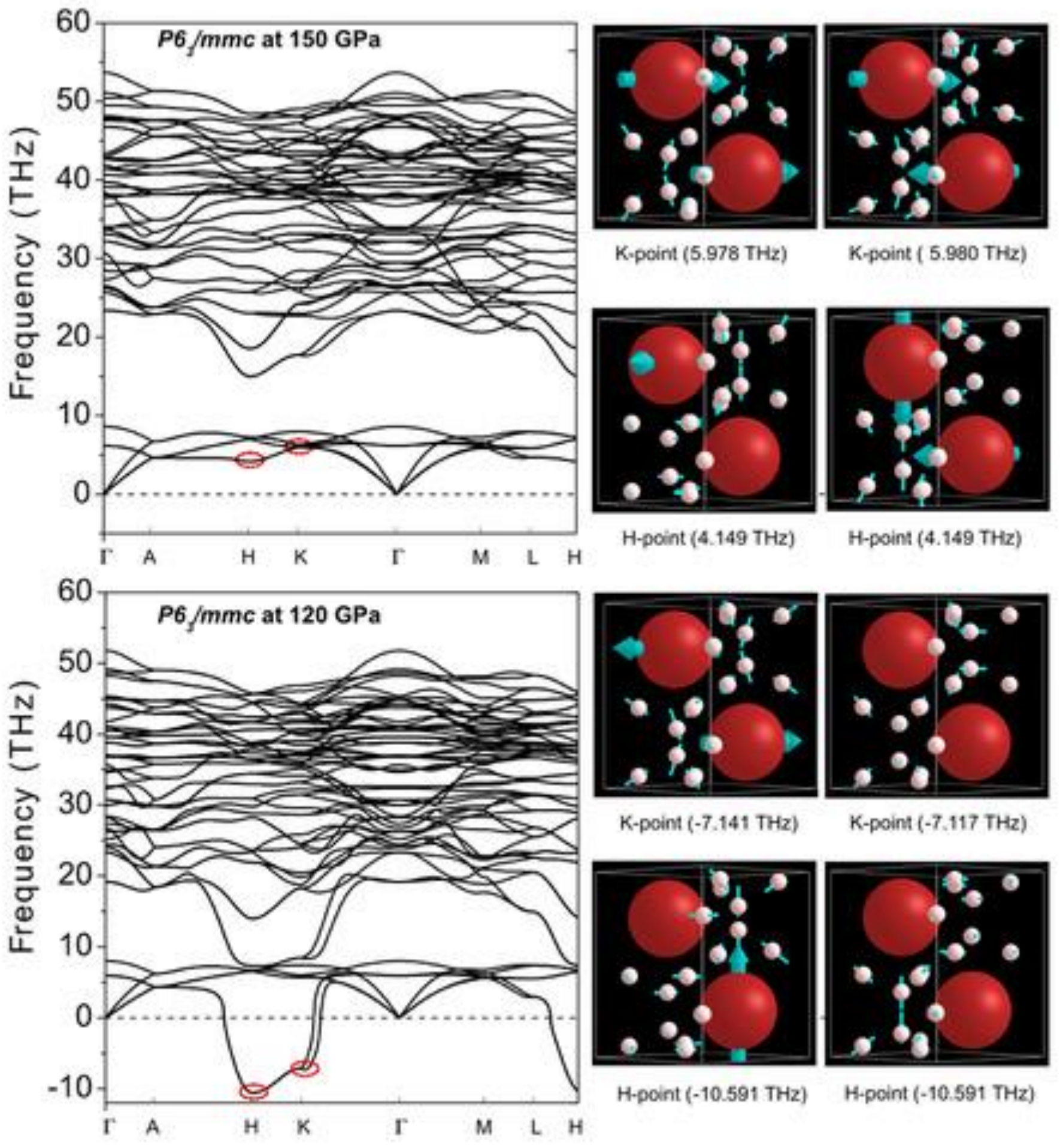

**Supplementary Figure S10: Phonon dispersion curves and selected mode displacements of the *P6₃/mmc*-CeH₉ at 120 and 150 GPa.** Unstable modes i.e., H- and K-point can be seen at 120 GPa. Large and small spheres show Ce and H atoms, respectively. Arrows refer to the displacement direction.

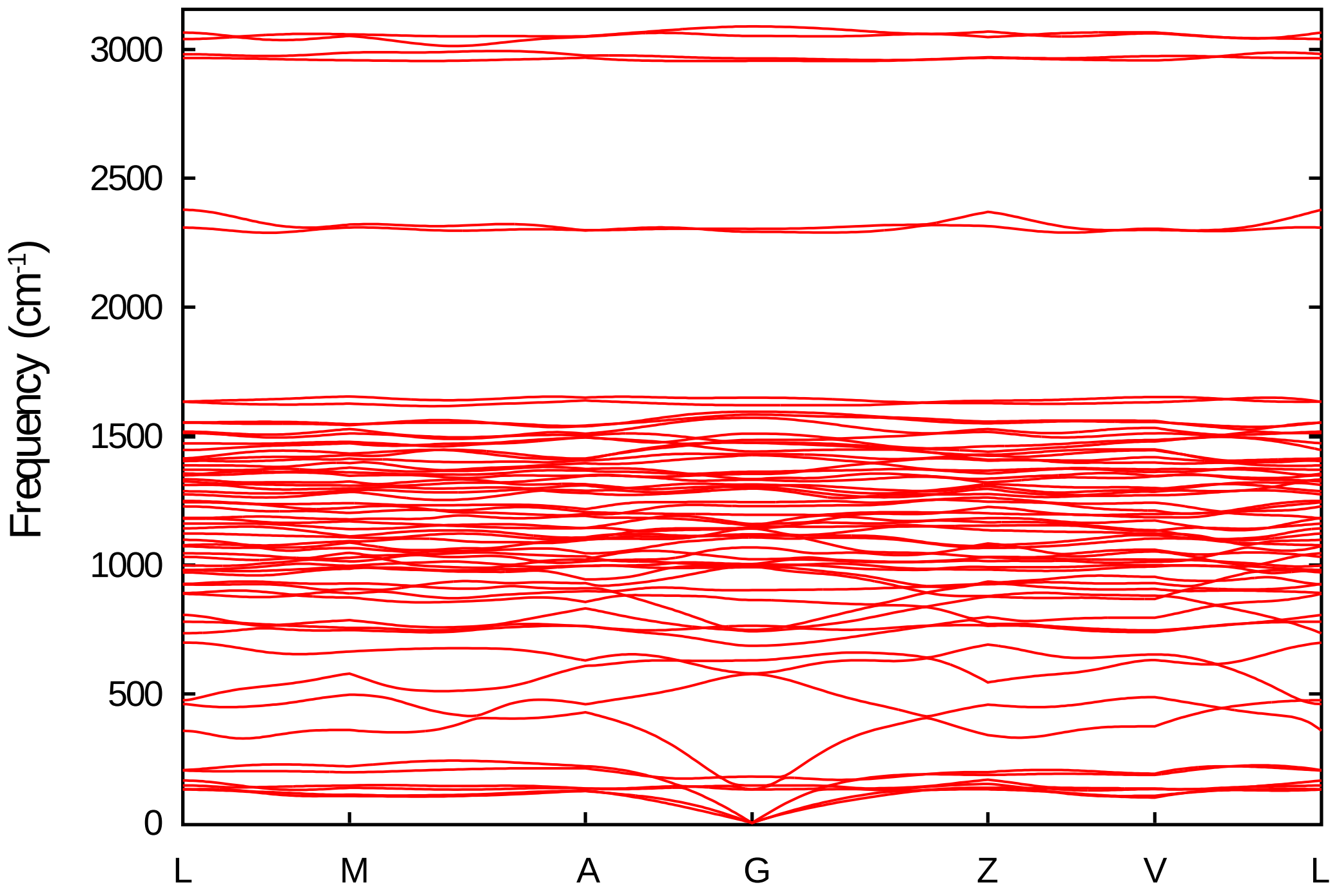

**Supplementary Figure S11: Calculated phonon dispersion curves of *C2/c*-$CeH_9$ at 100 GPa.** Phonon calculations show that lower symmetry structure *C2/c* does not have any imaginary frequency at 100 GPa, which simply refers to the stability of this phase. In addition to stability, lattice dynamics calculations show appearance of high frequency vibration of hydrogen atoms in the *C2/c* phase.

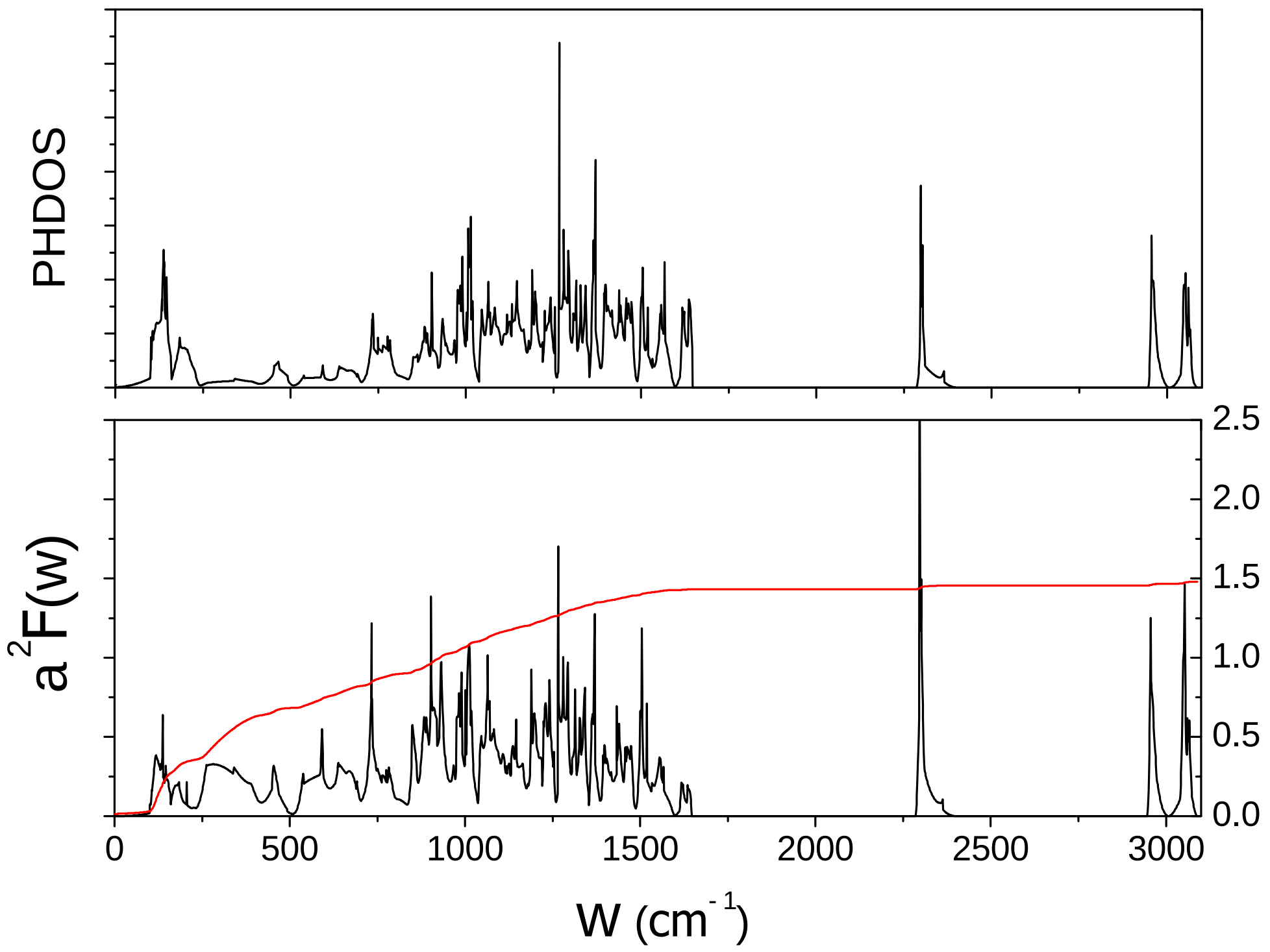

**Supplementary Figure S12: (a) Calculated phonon density of states (PHDOS), (b) Eliashberg EPC spectral functions $\alpha^2F(\omega)$, and electron- phonon integral $\lambda(\omega)$ (red line) of *C2/c*-$CeH_9$ 100 GPa.** Both phonon density of states and Eliashberg spectral function show a gap between medium-frequency and high-frequency vibration of H atoms. The electron-phonon coupling coefficient (red line) shows that medium-frequency H modes contribute the most to the electron-phonon coupling.

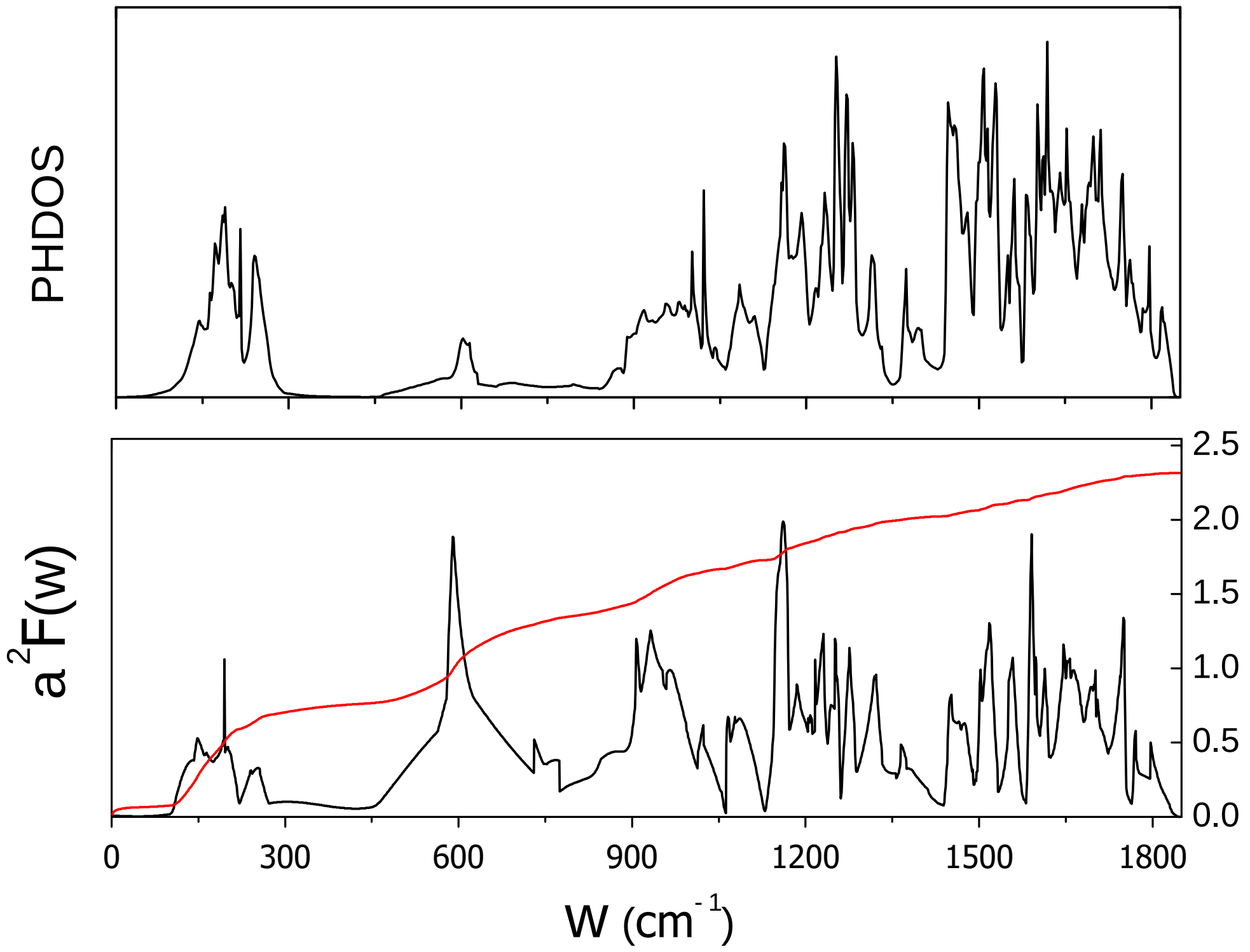

**Supplementary Figure S13: (a) Calculated phonon density of states (PHDOS), (b) Eliashberg EPC spectral functions $\alpha^2F(\omega)$, and electron- phonon integral $\lambda(\omega)$ (red line) of *$P6_3/mmc$*-$CeH_9$ 200 GPa.** Contrary to the low-symmetry phase of $CeH_9$, *$P6_3/mmc$* does not have hydrogen modes with frequency above 2000 cm$^{-1}$. The electron-phonon coupling coefficient (red line) shows that both medium-frequency and high-frequency H vibrations make a significant contribution to the electron-phonon coupling.

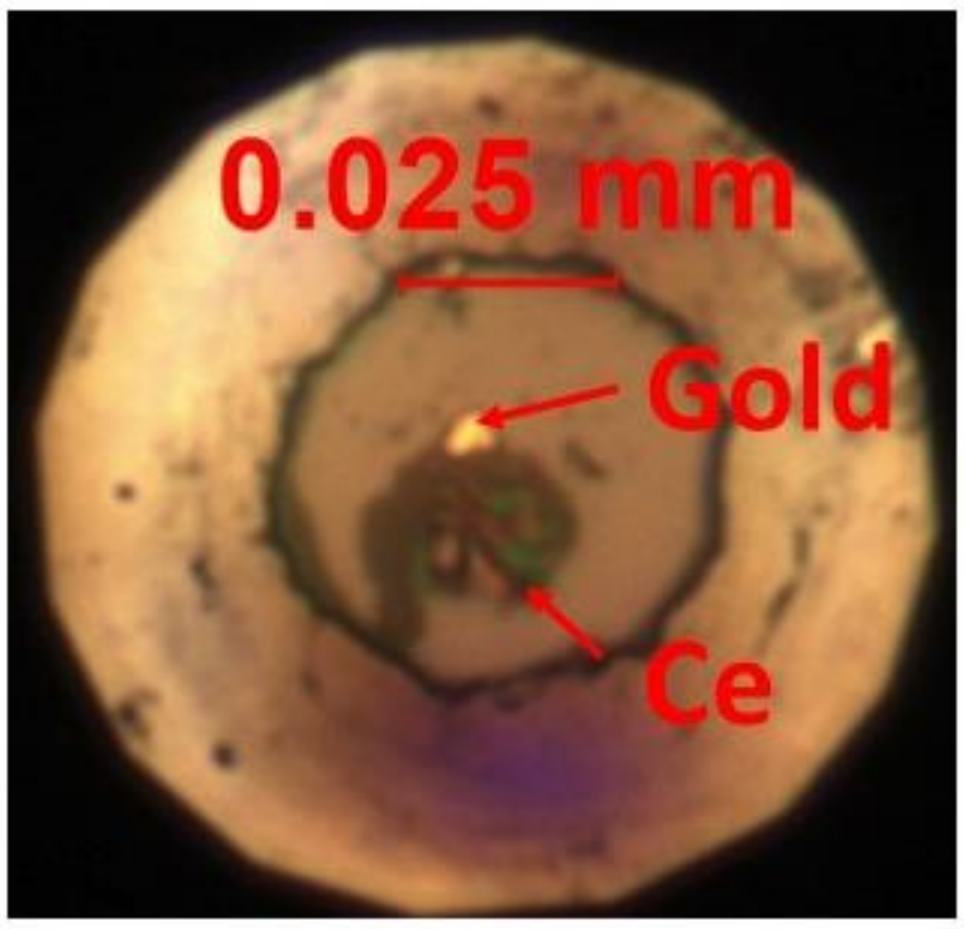

**Supplementary Figure S14: Image of sample loading at 9 GPa**. Image shows Ce (actually $CeH_2$) and Au surrounded by $H_2$ inside rhenium sample chamber at 9 GPa.

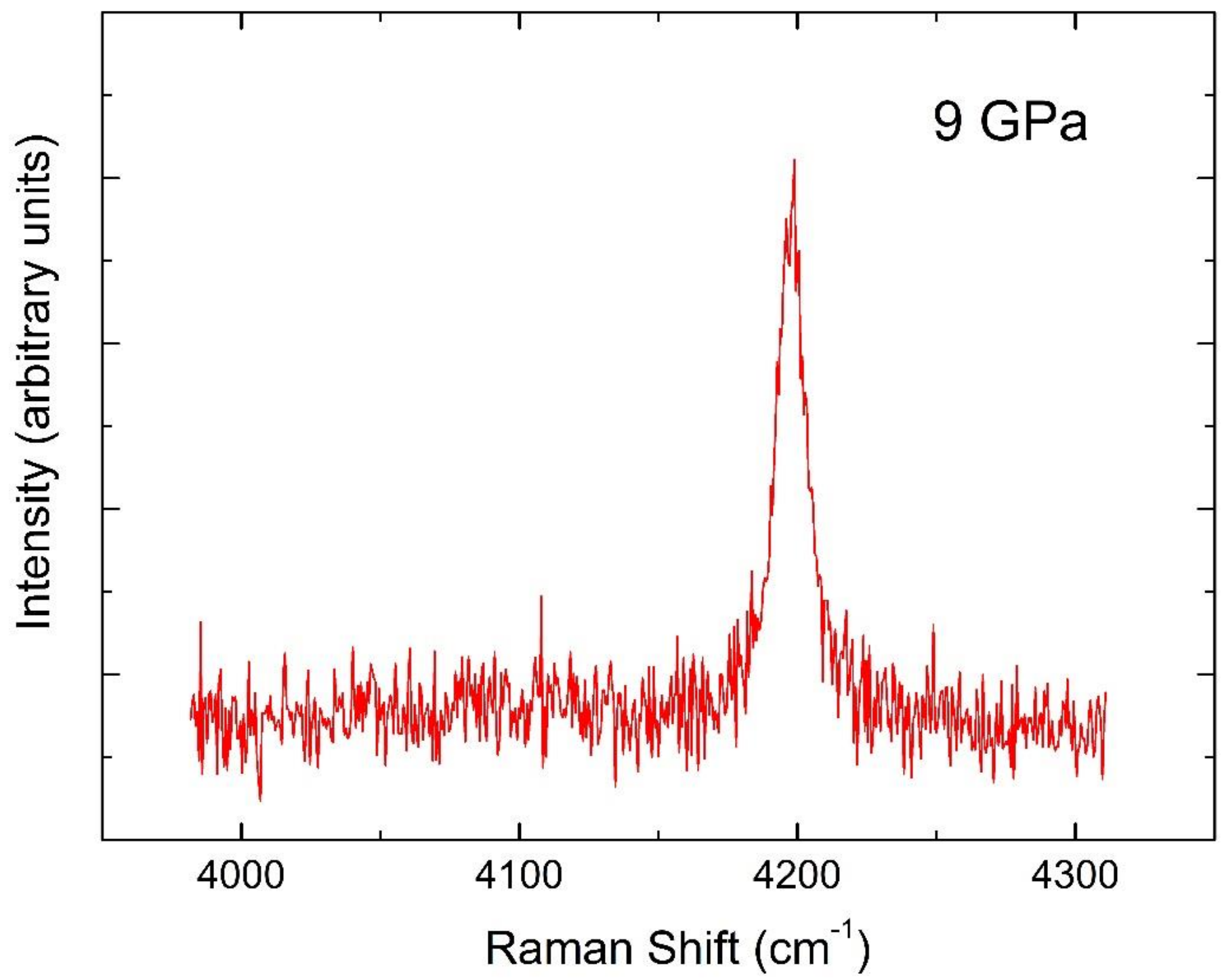

**Supplementary Figure S15: Raman spectra recorded from sample chamber at 9 GPa.** Raman spectra of $H_2$ vibron collected at 9 GPa from sample chamber confirming the presence of $H_2$ inside the sample chamber.

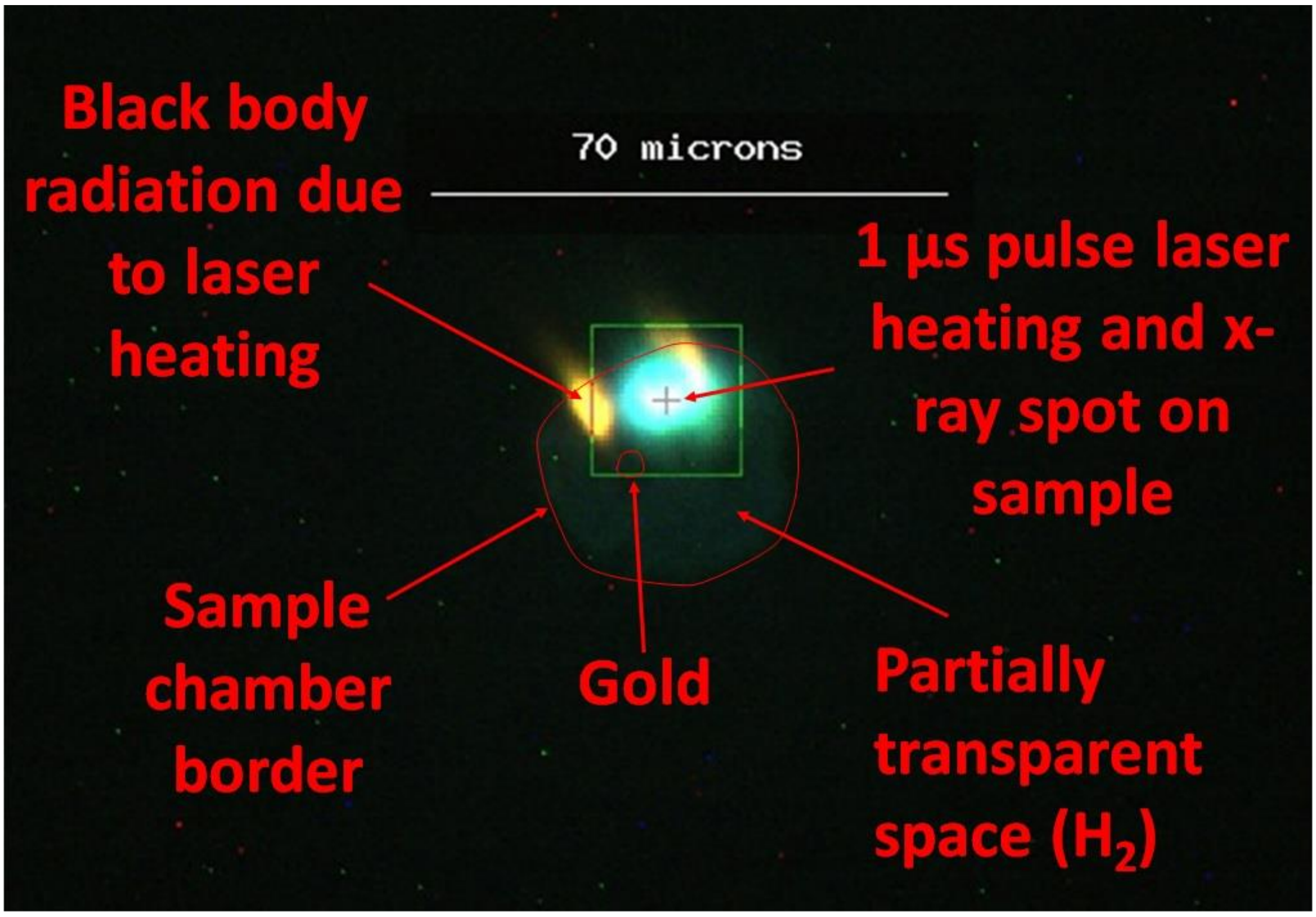

**Supplementary Figure S16: Image of *in-situ* ~1 µs pulse laser heating and x-ray on sample at 100 GPa recorded at beamline 13-IDD of GSECARS at APS.** Microsecond pulse laser heating of sample spot was always maintained at substantial distance from gasket corner (bright spot marked with an arrow) to avoid any unwanted reaction and contamination. Sample chamber edge and gold positions are marked with red line for clarity and labelled respectively in image. Partially transparent space shown with an arrow confirms the presence of hydrogen in sample chamber.

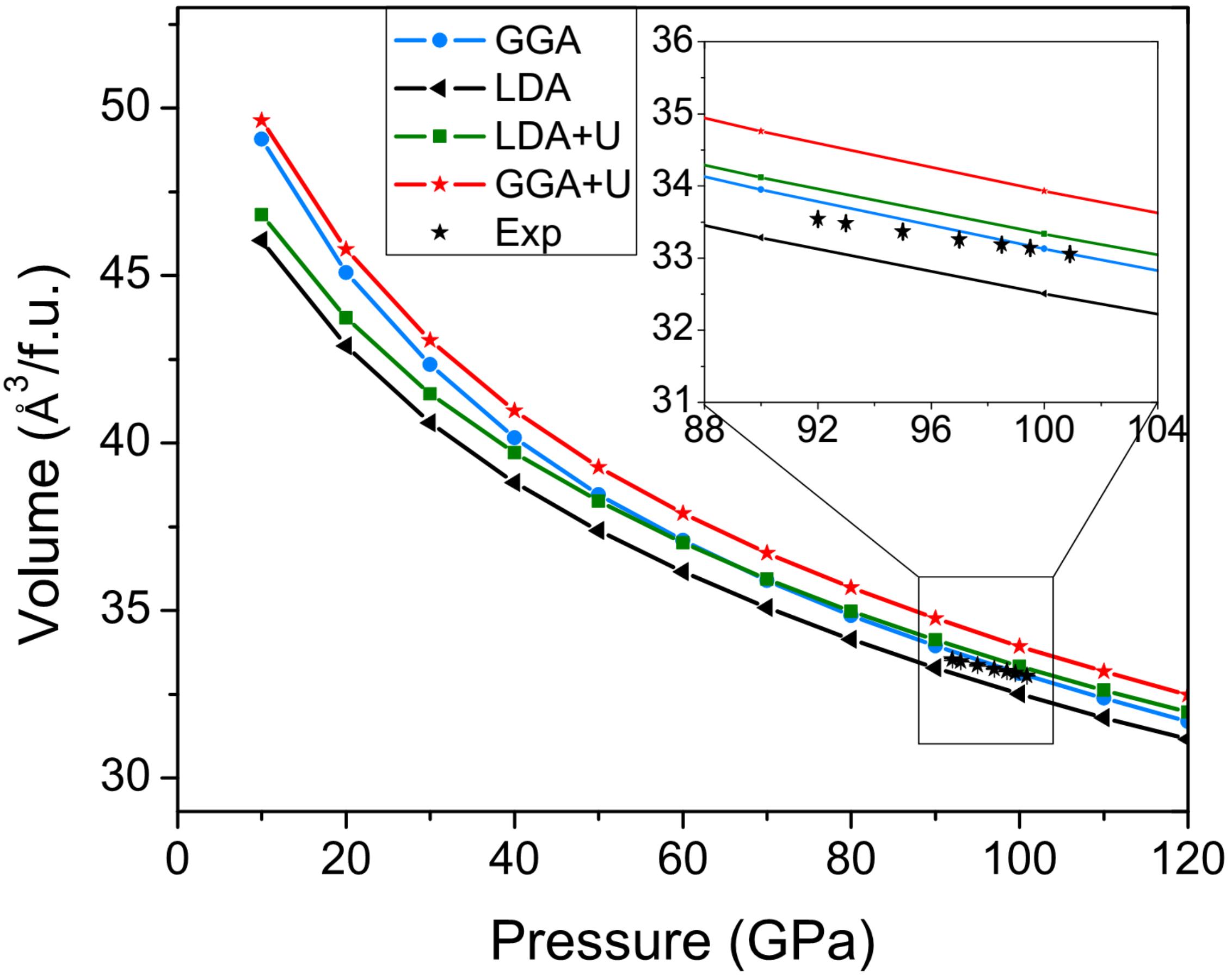

**Supplementary Figure S17: Comparison of P-V data of various theoretical methods with experimental data.**

**Table S1: Crystal structures parameters of the predicted Ce-H phases.**

| Phase | Space group | Lattice Parameters | Atoms | *x* | *y* | *z* |
|---|---|---|---|---|---|---|
| $CeH_{10}$ (200 GPa) | $Fm\bar{3}m$ | $a$ = 4.877 Å | Ce (4a) | 0.0000 | 0.0000 | 0.0000 |
| | | | H (8c) | 0.2500 | 0.2500 | 0.7500 |
| | | | H (32f) | 0.1227 | 0.1227 | 0.6227 |
| $CeH_9$ (200 GPa) | $P6_3/mmc$ | $a$ = 3.497 Å | Ce (2d) | 0.6667 | 0.3333 | 0.2500 |
| | | $c$ = 5.224 Å | H (12k) | 0.1550 | 0.8450 | 0.4351 |
| | | | H (4f) | 0.3333 | 0.6667 | 0.1464 |
| | | | H (2b) | 0.0000 | 0.0000 | 0.7500 |
| $CeH_9$ (100 GPa) | *C2/c* | $a$ = 3.685 Å | Ce (4e) | 0.5000 | 0.0823 | 0.2500 |
| | | $b$ = 6.427 Å | H (8f) | 0.9242 | 0.8255 | 0.4396 |
| | | $c$ = 6.652 Å | H (8f) | 0.5502 | 0.8258 | 0.0613 |
| | | $\beta$ = 123.60° | H (8f) | 0.8168 | 0.0862 | 0.0656 |
| | | | H (8f) | 0.1152 | 0.0840 | 0.8596 |
| | | | H (4e) | 0.0000 | 0.2482 | 0.2500 |
| $CeH_8$ (80 GPa) | $P6_3mc$ | $a$ = 3.727 Å | Ce (2b) | 0.6667 | 0.3333 | 0.7743 |
| | | $c$ = 5.576 Å | H (2b) | 0.6667 | 0.3333 | 0.1520 |
| | | | H (6c) | 0.1746 | 0.3492 | 0.9512 |
| | | | H (6c) | 0.8434 | 0.6869 | 0.0910 |
| | | | H (2a) | 0.0000 | 0.0000 | 0.2745 |
| $CeH_6$ (50 GPa) | $P6_3mc$ | $a$ = 3.704 Å | Ce (2b) | 0.6667 | 0.3333 | 0.2894 |
| | | $c$ = 5.733 Å | H (6c) | 0.1778 | 0.3556 | 0.1077 |
| | | | H (2a) | 0.0000 | 0.0000 | 0.4456 |
| | | | H (2a) | 0.0000 | 0.0000 | 0.7915 |
| | | | H (2b) | 0.3333 | 0.6667 | 0.4093 |
| $CeH_4$ (50 GPa) | *I4/mmm* | $a$ = 3.044 Å | Ce (2b) | 0.5000 | 0.5000 | 0.0000 |
| | | $c$ = 6.021 Å | H (4e) | 0.0000 | 0.0000 | 0.8657 |
| | | | H (4d) | 0.0000 | 0.5000 | 0.2500 |
| $CeH_3$ (50 GPa) | $Pm\bar{3}n$ | $a$ = 3.815 Å | Ce (2a) | 0.5000 | 0.5000 | 0.5000 |
| | | | H (6d) | 0.2500 | 0.5000 | 0.0000 |
| $CeH_3$ (0 GPa) | $Pm\bar{3}n$ ($\beta$-$UH_3$-type) | $a$ = 6.570 Å | Ce (2a) | 0.0000 | 0.0000 | 0.0000 |
| | | | Ce (6c) | 0.2500 | 0. 0000 | 0.5000 |
| | | | H (24k) | 0.0000 | 0.1552 | 0.6953 |

**Table S2: Comparison of electronic density of states of various hydrides**

| Hydride | Pressure (GPa) | N(Ef) (states/eV/f.u.) |
|---|---|---|
| $P6_3/mmc$-$CeH_9$ | 100 GPa | 0.918 |
| $P6_3/mmc$-$CeH_9$ | 150 GPa | 0.812 |
| $P6_3/mmc$-$CeH_9$ | 200 GPa | 0.732 |
| $Im\bar{3}m$-$H_3S$ | 200 GPa | 0.525 |
| $Fm\bar{3}m$ -$LaH_{10}$ | 200 GPa | 0.735 |

**Table S3: Comparison of electron-phonon coupling coefficient (λ), logarithmic average phonon frequency ($\omega_{log}$) and $T_c$ for various hydrides. ([a])**

| Hydride | Space group | λ | $\omega_{log}$ (K) | $T_c$ (K) | Pressure (GPa) | reference |
|---|---|---|---|---|---|---|
| $CeH_9$ | $P6_3/mmc$ | 2.30 | 740 | 117 | 200 | This study |
| $CeH_9$ | $C2/c$ | 1.48 | 662 | 75 | 100 | This study |
| $H_3S$ | $Im\bar{3}m$ | 2.19 | 1335 | 203 | 200 | 4 |
| $YH_{10}$ | $Im\bar{3}m$ | 2.58 | 1282 | 326 | 250 | 5 |
| $LaH_8$ | $C2/m$ | 1.12 | 1591 | 131 | 300 | 5 |
| $LaH_{10}$ | $Fm\bar{3}m$ | 3.41 | 848 | 238 | 210 | 5 |
| $UH_7$ | $P6_3/mmc$ | 0.83 | 873.8 | 47.6 | 20 | 6 |
| | | 0.95 | 764.9 | 57.5 | 0 | |
| $UH_8$ | $Fm\bar{3}m$ | 0.73 | 873.7 | 27.5 | 50 | 6 |
| | | 1.13 | 450.3 | 37.6 | 0 | |
| $UH_9$ | $P6_3/mmc$ | 0.67 | 933.4 | 35.8 | 300 | 6 |
| $AcH_{10}$ | $R\bar{3}m$ | 3.46 | 710.9 | 204.1 | 200 | 7 |
| $AcH_{16}$ | $P\bar{6}m2$ | 2.16 | 1054 | 199.2 | 150 | 7 |
| $ThH_{10}$ | $Fm\bar{3}m$ | 2.19 | 1042.8 | 194.4 | 100 | 8 |

**Table S4: Comparison of lattice parameters and volume of *$P6_3/mmc$*-$CeH_9$ at 100 GPa by various DFT functionals**

| **Functional** | ***a*** **(Å)** | ***c*** **(Å)** | **Vol ($Å^3$/f.u.)** |
|---|---|---|---|
| LDA | 3.6827 | 5.5356 | 32.510 |
| LDA+U (6 eV) | 3.7072 | 5.6018 | 33.335 |
| GGA | 3.7022 | 5.5813 | 33.125 |
| GGA+U (6 eV) | 3.7259 | 5.6445 | 33.930 |
| Experiment | 3.7110 | 5.5429 | 33.053 |

**Table S5: Comparison of bulk modulus of *$P6_3/mmc$*-$CeH_9$ by various DFT functionals**

| Functional | bulk modulus $K0$ (GPa) | $V_0$ ($A^3$/f.u.) | $K_0'$ |
|---|---|---|---|
| LDA | 98.1 | 49.9 | 4.0 |
| LDA+U (6 eV) | 105.4 | 50.4 | 4.0 |
| GGA | 80.5 | 53.4 | 4.0 |
| GGA+U (6 eV) | 84.4 | 54.0 | 4.0 |

## References

1 Avisar, D. & Livneh, T. Raman scattering by phonons and crystal-field excitations in cerium hydrides. *Journal of Alloys and Compounds* **494**, 11-16, doi:https://doi.org/10.1016/j.jallcom.2009.11.108 (2010).

2 Jacobsen, M. K., Velisavljevic, N., Dattelbaum, D. M., Chellappa, R. S. & Park, C. High pressure and temperature equation of state and spectroscopic study of CeO 2. *Journal of Physics: Condensed Matter* **28**, 155401 (2016).

3 Korst, W. L. & Warf, J. C. Rare Earth-Hydrogen Systems. I. Structural and Thermodynamic Properties. *Inorganic Chemistry* **5**, 1719-1726, doi:10.1021/ic50044a018 (1966).

4 Duan, D. *et al.* Pressure-induced metallization of dense (H2S)2H2 with high-Tc superconductivity. *Scientific Reports* **4**, 6968, doi:10.1038/srep06968 (2014).

5 Liu, H., Naumov, I. I., Hoffmann, R., Ashcroft, N. W. & Hemley, R. J. Potential high Tc superconducting lanthanum and yttrium hydrides at high pressure. *Proceedings of the National Academy of Sciences* **114**, 6990-6995, doi:10.1073/pnas.1704505114 (2017).

6 Kruglov, I. A. *et al.* Uranium polyhydrides at moderate pressures: prediction, synthesis, and expected superconductivity. *arXiv:1708.05251* (2017).

7 Semenok, D. V., Kvashnin, A. G., Kruglov, I. A. & Oganov, A. R. Actinium Hydrides AcH10, AcH12, and AcH16 as High-Temperature Conventional Superconductors. *The Journal of Physical Chemistry Letters*, 1920-1926, doi:10.1021/acs.jpclett.8b00615 (2018).

8 Kvashnin, A. G., Semenok, D. V., Kruglov, I. A. & Oganov, A. R. High-Temperature Superconductivity in Th-H System at Pressure Conditions. *arXiv:1711.00278* (2017).